\documentclass{ws-procs9x6}
\usepackage{epsfig}

\def\nabstar#1{\nabla\kern-0.5pt\smash{\raise 4.5pt\hbox{$\ast$}}
               \kern-4.5pt_{#1}}

\def\Nf{N_{\rm f}}
\def\psibar{\bar{\psi}}

\def\mc{m_{\rm cr}}
\def\muq{\mu_{\rm q}}

\begin{document}

\title{Lattice QCD with a chiral twist\footnote{based on lectures given at the 
school ``\uppercase{P}erspectives in \uppercase{L}attice \uppercase{G}auge 
\uppercase{T}heories'', \uppercase{N}ara, \uppercase{J}apan, \uppercase{O}ctober 31-
\uppercase{N}ovember 14, 2005}}

\author{S. SINT}

\address{School of Mathematics,\\
Trinity College Dublin,\\
Dublin 2, Ireland\\
{\tt sint@maths.tcd.ie}}

\maketitle

\abstracts{In these lectures I explain how chiral symmetry
of continuum QCD naturally leads to a class of lattice regularisations 
known as  twisted mass QCD (tmQCD). As compared to standard
Wilson quarks, its advantages are the absence of unphysical zero modes, 
the possibility to circumvent lattice renormalisation problems 
and automatic O($a$) improvement. On the other hand, the physical 
parity and flavour symmetries are explicitly broken. 
I discuss these aspects and then turn to the theory in a finite 
space-time volume with Schr\"odinger functional boundary conditions.
Again, chiral transformations  of the continuum theory may be used
as a guide to formulate an alternative lattice regularisation 
of the Schr\"odinger functional, with interesting 
applications to renormalization problems in QCD.}

\section{Introduction}

In recent years, twisted mass QCD (tmQCD) has become a popular
variant of lattice QCD with Wilson-type 
quarks~\cite{Frezzotti:1999vv,Frezzotti:2000nk,Frezzotti:2001ea}. Initially designed
to render the (partially) quenched approximation well-defined through 
the  elimination of unphysical zero modes, it was soon
realised that tmQCD could also be used to circumvent some notorious 
lattice renormalization problems~\cite{Frezzotti:1999vv,Frezzotti:2000nk}. 
Later, Frezzotti and Rossi~\cite{Frezzotti:2003ni} observed that
scaling violations in tmQCD can be reduced to
O($a^2$) without the need for all the O($a$) counterterms required
with standard Wilson quarks ($a$ being the lattice spacing).
This property, referred to as ``automatic O($a$) improvement",
has attracted further attention and a number of groups
have started large scale numerical simulations using tmQCD.
In these lectures I do not attempt to review this work in 
progress\footnote{See \cite{Shindler:2005vj} for a review and further
references.}. Here I would rather like to give an introduction to the
basic concepts. This includes in particular a discussion of O($a$) improvement 
and the question whether it is  compromised by currently
used non-perturbative renormalization procedures based on the 
QCD Schr\"odinger functional
(SF-schemes). In fact, the standard Schr\"odinger functional
boundary conditions turn out to be difficult to reconcile with 
automatic O($a$) improvement and the
construction of an alternative set-up for the Schr\"odinger functional
may therefore be advantageous.

This writeup is organised as follows: I start with the interplay
between the choice of the quark mass term and the form taken by
parity, flavour and chiral symmetry transformations (sect.~2). 
After a reminder of standard Wilson quarks and the problem 
of unphysical zero modes (section~3), lattice tmQCD is introduced
in sect.~4. Based on the formal continuum theory a dictionary between
tmQCD and QCD correlation functions is readily established, which is expected
to hold between properly renormalised correlation functions.
It then becomes clear how to by-pass certain renormalization problems of standard Wilson
quarks (sect.~5), and the computation of $B_K$ is discussed in some detail.
In sect.~6 automatic O($a$) improvement of tmQCD is analysed using Symanzik's 
effective theory. Potential problems of tmQCD associated with 
flavour and parity breaking are shortly mentioned in sect.~7.
In sect.~8, the properties of Schr\"odinger functional 
renormalisation schemes (SF schemes) are discussed.
Motivated by the clash of the standard set-up with automatic O($a$) improvement
and by the slow decoupling of heavy quarks in mass-dependent SF schemes, 
a modified definition of the Schr\"odinger functional is proposed,
and its effectiveness regarding O($a$) improvement
is illustrated in an example taken from perturbation theory.
Section~9 contains some conclusions.

\section{Continuum QCD and chiral transformations}

Let us consider the continuum action of QCD with $\Nf=2$ massless 
quarks\footnote{Conventions used for 
Euclidean $\gamma$-matrices in 4 dimensions:
$\{\gamma_\mu,\gamma_\nu\}=2\delta_{\mu\nu}$, $\gamma_\mu^\dagger=\gamma_\mu$,
where $\mu,\nu=0,1,2,3$, and $\gamma_5=\gamma_0\gamma_1\gamma_2\gamma_3$, 
$\sigma_{\mu\nu}=\frac{i}{2}[\gamma_\mu,\gamma_\nu]$.}.
Decomposing the action into a pure gauge and a fermionic part,
$S=S_g+S_f$, we here focus on the fermionic part,
\begin{eqnarray} 
 S_f &=& \int{\rm d}^4 x\,\, \psibar(x)\mbox{$D${\hspace{-2.1mm}/}}\psi(x), 
  \qquad
 \mbox{$D${\hspace{-2.1mm}/}}=\gamma_{\mu}D_{\mu}.
\end{eqnarray}
The quark and antiquark fields $\psi,\psibar$ are flavour doublets,
interacting minimally with the gluon field $A_\mu$ 
via the covariant derivative $D_\mu = \partial_\mu+A_\mu$. 
The massless fermionic action has a global chiral-flavour SU(2)$\times$SU(2)
invariance, corresponding to the transformations,
\begin{eqnarray}
 \psi \rightarrow \psi^{\prime} &=&
  \exp(i\omega^a_V\tau^a/2)\exp(i\omega^b_A\gamma_5\tau^b/2)\psi, \nonumber\\
 \bar\psi \rightarrow \bar\psi^{\prime} &=&
  \bar\psi\exp(i\omega^b_A\gamma_5\tau^b/2)\exp(-i\omega^a_V\tau^a/2), 
 \label{eq:su2su2trafo}
\end{eqnarray}
where $\tau^a$  $(a=1,2,3)$ are Pauli matrices, $\omega^a_{V,A}$ are
transformation parameters\footnote{Summation over repeated indices $a,b=1,2,3$
is understood.}. This notation distinguishes
the axial from the vector generators (corresponding to the flavour or isospin 
SU(2) subgroup) in a standard way. 

A quark mass term breaks the chiral flavour symmetry explicitly,
leaving only the vector or isospin symmetry intact.
The above notation for the symmetry transformations
was introduced with the standard quark mass term $\bar\psi\psi$
in mind, but e.g.~the choice
\begin{eqnarray}
 \psibar^{\prime}\psi^{\prime} 
= \psibar\exp(i\omega^a_A\gamma_5\tau^a)\psi
= \cos(\omega_A)\psibar\psi + i\sin(\omega_A)u^a_A \psibar\gamma_5\tau^a\psi,
\label{eq:4}
\end{eqnarray}
would be completely equivalent. Here, $\omega_A$ denotes the modulus of
$(\omega^1_A,\omega^2_A,\omega^3_A)$ and $u^a_A = \omega^a_A/\omega_A$
is a unit vector. 
In fact, it is only after the introduction of the quark mass
term that the distinction between axial and vector symmetries
acquires a meaning. By definition, the vector symmetry transformations
are those which leave the quark mass term invariant. Similarly, the quark
mass term is supposed to be invariant under parity transformations.
As a consequence, the form of a symmetry transformation
depends on the choice of the mass term. While a standard mass term
implies that a parity transformation can be realised as 
\begin{eqnarray}
\psi(x_0,{\bf x})\rightarrow\gamma_0\psi(x_0,-{\bf x}),\qquad
\psibar(x_0,{\bf x})\rightarrow\psibar(x_0,-{\bf x})\gamma_0,
\end{eqnarray}
the alternative choice of (\ref{eq:4}) for the mass term means that
a parity transformation will look more complicated, for instance
\begin{eqnarray}
 \psi(x_0,{\bf x}) & \rightarrow &
  \gamma_0\exp(i\omega^a_A\gamma_5\tau^a)\psi(x_0,-{\bf x}), \nonumber\\
 \bar\psi(x_0,{\bf x}) & \rightarrow &
  \bar\psi(x_0,-{\bf x})
 \exp(i\omega^a_A\gamma_5\tau^a)\gamma_0.
\end{eqnarray}
Similarly, the isospin transformation obtained with a standard 
mass term corresponds to (\ref{eq:su2su2trafo}) with
all axial transformation parameters set to zero, 
$\omega^a_A =0$ (whence the notation), whereas the mass term (\ref{eq:4}) leads to 
the much less intuitive formula
\begin{eqnarray}
 \psi & \rightarrow &
  \exp(-i\omega^a_A\gamma_5\tau^a/2)\exp(i\omega^b_V\tau^b/2)
  \exp(i\omega^c_A\gamma_5\tau^c/2)\psi, \nonumber\\
 \bar\psi & \rightarrow &
  \bar\psi\exp(i\omega^a_A\gamma_5\tau^a/2)\exp(-i\omega^b_V\tau^b/2)
  \exp(-i\omega^c_A\gamma_5\tau^c/2),
\end{eqnarray}
where $\omega^b_V$ $(b=1,2,3)$ are transformation parameters 
while $\omega^a_A$ $(a=1,2,3)$
are again fixed. The situation is reminiscent of the choice of a coordinate system, and
our intuition about the form of symmetry transformations is thus based on
a particular choice of ``field coordinates''. 
Of course, this raises the question why one should 
deviate from the standard choice of the mass term.
In the continuum and for regularisations preserving chiral symmetry
there is indeed no point in introducing a twisted mass term, for 
any non-standard choice could be brought into the standard form by using an axial rotation,
which, being a symmetry of the massless theory, has no further effects.
The situation is different in regularisations which
break chiral symmetry, such as lattice regularisations with 
Wilson type quarks. One may then obtain different regularisations of QCD which
have equivalent continuum limits but differ at the cutoff level. 
This will be made more precise a bit later. 

\section{Standard Wilson quarks}

Standard Wilson quarks are characterised by the fermionic lattice action,
\begin{eqnarray}
 S_f &=& a^4\sum_x \bar\psi(x)(D_{\rm W}+m_0)\psi(x), \\
 D_{\rm W} &=& \sum_{\mu=0}^3
  \left\{{\textstyle\frac{1}{2}}(\nabla_{\mu}^{}+\nabla_{\mu}^{\ast})\gamma_{\mu}
    -a\nabla_{\mu}^{\ast}\nabla_{\mu}^{}\right\}.
\label{eq:Dw}
\end{eqnarray}
Here, $m_0$ is a bare mass parameter and the covariant lattice derivatives 
in the Wilson-Dirac operator are defined as usual 
(see ref.~\cite{Luscher:1996sc} for unexplained notation). Assuming $\Nf$ quark flavours
the lattice action has an exact U($\Nf$) vector symmetry, and 
is invariant under axis permutations, reflections such as parity and charge conjugation. 
Furthermore, unitarity of lattice QCD with Wilson quarks has
been rigorously established~\cite{Luscher:1976ms}. These nice properties of
standard Wilson quarks come with a price: all axial symmetries are explicitly broken
by the last term in eq.~(\ref{eq:Dw}), called the Wilson term.
This has a number of consequences:
\begin{enumerate}
\item Linear mass divergence:
the quark mass term is not protected against
additive renormalization, i.e.~any renormalized quark mass is of the 
form 
$
 m_{\rm R} = Z_m (m_0-\mc),
$
where the critical mass is linearly divergent, i.e.~$\mc \propto 1/a$.
\item Axial current renormalization:
since axial transformations are not an exact symmetry, there 
is no exact current algebra, and the non-singlet axial
current requires a non-trivial multiplicative renormalization
to restore current algebra up to O($a$) effects.
\item
Definition of the chiral condensate as expectation value of a local operator:
the renormalised iso-singlet scalar density has the structure,
\begin{equation}
 (\psibar\psi)_{\rm R} = Z_{\rm S^0}\{\psibar\psi + c_{\rm S} a^{-3}\}.
 \label{eq:cubic}
\end{equation}
In a regularisation which respects chiral symmetry, the additive renormalization
constant $c_{\rm S}$ would be proportional to $am$, with $m$ being a multiplicatively
renormalisable bare quark mass. This means that the chiral condensate is well-defined
in the chiral limit once its {\em multiplicative} renormalisation has been carried out.
In contrast, with Wilson quarks one first needs to subtract the cubic power divergence,
even in the chiral limit.
\item Cutoff effects:
the leading cutoff effects with Wilson-type fermions are proportional to $a$, rather than $a^2$.
Again, this is a consequence of chiral symmetry breaking. This is easily seen
by looking at the structure of the counterterms which are to be included
for the on-shell O($a$) improvement of the theory \`a la Symanzik~\cite{Luscher:1996sc}.
\end{enumerate}
From a field theoretical point of view this illustrates
the proliferation of additional counterterms in a case where
the regularisation breaks a continuum symmetry. One should note, however,
that there is no remaining theoretical or conceptual problem.

\subsection{Wilson quarks and unphysical fermionic zero modes}

Nevertheless, technical problems may arise within the current practice
of numerical simulations with Wilson-type quarks. This is related to
the fact that, for a given gauge background field, the massive
Wilson-Dirac operator $D_W+m_0$ is not protected against 
zero modes unless the bare mass parameter $m_0$ is positive.
However, due to additive quark mass renormalisation,
the masses of the light quarks typically correspond to 
negative bare mass parameters, which leaves the Wilson-Dirac operator 
unprotected against zero modes in the physically interesting region.
These modes are considered unphysical, since one expects from the 
continuum theory that any non-zero value of the renormalised quark mass prohibits 
zero modes of the Dirac operator.

It is instructive to look at a typical 
fermionic correlation function, such as the pion propagator given by
\begin{eqnarray} 
 G^{ab}(x-y) &=& -\left\langle\psibar(x)\gamma_5\frac{\tau^a}{2}\psi(x)
                              \psibar(y)\gamma_5\frac{\tau^b}{2}\psi(y)
                  \right\rangle\nonumber\\
	     &=&  -{\mathcal{Z}}^{-1}\int D[U,\psi,\psibar]
  {\rm e}^{-S}\psibar(x)\gamma_5\frac{\tau^a}{2}\psi(x)
                              \psibar(y)\gamma_5\frac{\tau^b}{2}\psi(y),
\end{eqnarray}
where $\tau^a$, $a=1,2,3$,  are the Pauli matrices acting in 
flavour space and ${\mathcal{Z}}=\langle 1 \rangle$.
It is convenient to introduce the operator,
\begin{equation}
  Q=\gamma_5(D_{\rm W}+m_0),\qquad Q = Q^\dagger,
\end{equation}
which acts in single flavour space.
Integrating over the quark and anti-quark fields one obtains
\begin{equation} 
G^{ab}(x-y)=\frac{1}{2}\delta^{ab}{\mathcal{Z}}^{-1}\int D[U] {\rm e}^{-S_g}
               \det\left(Q^2\right)  {\rm tr}\left[Q^{-1}(x,y)Q^{-1}(x,y)\right],
\end{equation}
where the flavour structure has been reduced analytically and the remaining
trace is over colour and spin indices. The important point to notice is that
the resulting expression is never singular. Denoting the eigenfunction
of $Q$ for a given eigenvalue $\lambda_i$ by $\varphi_i(x)$, 
the pion propagator takes the form
\begin{eqnarray}
 G^{ab}(x-y)&=&\frac{1}{2}\delta^{ab}{\mathcal{Z}}^{-1}\int D[U] {\rm e}^{-S_g}
  \left(\prod_{i}\lambda_i^2\right)\nonumber\\
  &&\hphantom{123}\times \sum_{j,k}\lambda_j^{-1}\lambda_k^{-1}
  \varphi_j(x)\varphi^\ast_j(y)\varphi_k(x)\varphi_k^\ast(y).
\end{eqnarray}
In other words, the eigenvalues in the denominator are 
always compensated by corresponding factors from the determinant. 
The limit of vanishing eigenvalues is always regular and
a strict lower bound on the eigenvalue spectrum is not required 
for the theory to be well-defined.

However, the absence of a lower bound on $|\lambda_i|$ may still 
lead to technical problems, either due to the use of 
unphysical approximations or due to the set-up of numerical simulations:

\subsubsection{Quenched and partially quenched approximations}

\begin{figure}[ht]
\epsfxsize=8cm   
\epsfbox{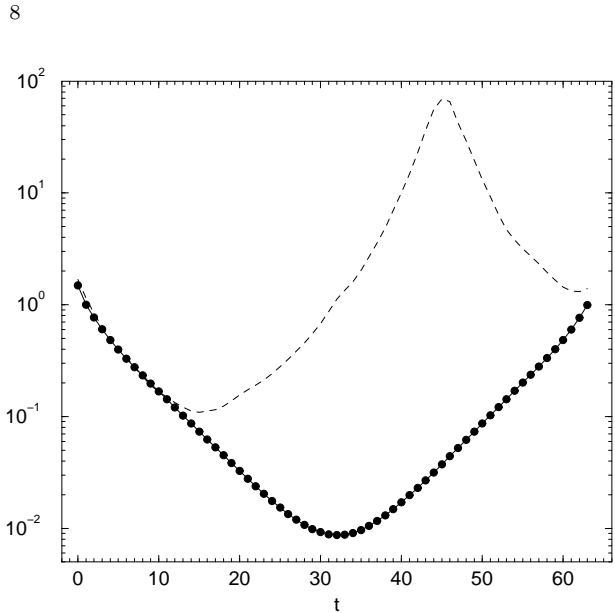}
\caption{The pion propagator vs. time separation
from a quenched simulation on a $32^3\times 64$ lattice at $\beta=6.2$
(cf.~text for an explanation).\label{fig:exceptional}}
\end{figure}

As the computational cost for the generation of a an ensemble of gauge field
configurations is dominated by the inclusion of the quark determinant,
a widely used approximation consists in omitting the determinant when taking
the average over gauge fields. The quark propagators with the eigenvalues in the
denominator may then become singular, and gauge field configurations where this
happens are called ``exceptional". The example in figure~\ref{fig:exceptional} 
taken from~\cite{Schierholz:1998bq} shows the ensemble average of the pion propagator 
over all gauge configurations but a a single exceptional one (dashed line), 
where the propagator deviates dramatically from the average (dots and solid line). 
The inclusion of the exceptional configuration in the 
ensemble average would lead to much larger errors, while its omission 
invalidates the Monte Carlo procedure. In principle one should say 
that the quenched approximation with Wilson type quarks is ill-defined, 
since zero modes are bound to occur if the ensemble of gauge configurations is large 
enough. However, the frequency of near zero modes depends very sensitively 
upon the bare quark mass and is in fact a function of the lattice size and 
all the other bare parameters in the lattice action. 
One may therefore think of the quenched approximation as being operationally defined, 
if for an ensemble of, say, a few hundred configurations the problem is typically absent. 
``Safe" parameter ranges may then be quoted for a given action, but this situation 
is clearly unsatisfactory. In particular, as the problem is not sharply defined, 
one may always be unlucky and encounter near zero modes even at parameter values 
which have previously been considered safe. In practice it is this problem 
which has limited the approach to the chiral limit, 
rather than finite volume effects due to the pions becoming too light.

Obviously, the problem is expected to disappear once the quark determinant 
is properly included. Usually this is done by including the complete determinant 
in the effective gauge field measure used in the importance sampling, 
and the probability for a gauge configuration to be included 
in the ensemble becomes proportional to the eigenvalues. 
Exceptional configurations are then never produced. However, even in this case, 
one is often interested in varying the valence quark masses independently
of the sea quark masses, a situation which is referred to as 
the partially quenched approximation. One may also have different numbers of 
valence and sea quarks, or Wilson valence quarks and sea quarks of a different kind. 
In all these cases one expects similar problems with unphysical zero modes 
as in the quenched approximation.

\subsubsection{Potential problems in the Hybrid Monte Carlo algorithm}

Most numerical simulations use some variant of the Hybrid Monte Carlo 
algorithm~\cite{Duane:1987de}. 
Integrating the molecular dynamics trajectories in fictitious phase space 
then requires the evaluation of the fermionic force term and 
thus the inversion of the Dirac operator at each step in 
molecular dynamics time. The force term may become very large if an 
exceptional configuration  is encountered, and the molecular dynamics 
integrator tends to become unstable if the product of the 
force and the step size exceeds a certain critical value~\cite{Joo:2000dh}.
To avoid this situation one may hence be forced to decrease the step size 
to very small values thereby increasing the cost of the simulation.
It is likely that this problem was at the
heart of the difficulties encountered in the past with simulations of 
Wilson type quarks~\cite{Bernard:2002pd}.
However, various developments over the past few years seem to have 
solved this problem (see~\cite{Giusti} for a recent account of current 
simulation algorithms and cost estimates and \cite{Kennedy:2006ax} for further discussion).

\section{Twisted mass lattice QCD}

Initially the main motivation for introducing a twisted mass term was the problem 
with zero modes discussed in the previous section.
The lattice action for a doublet $\psi$ of $\Nf=2$ mass degenerate
quarks is now given by
\begin{eqnarray}
 S_f &=& a^4\sum_x \bar\psi(x)(D_{\rm W}+m_0+i\muq\gamma_5\tau^3)\psi(x),
\end{eqnarray}
where $\muq$  denotes the bare twisted mass parameter. It is easy to see that the
presence of this parameter elimiates any unphysical zero modes, for
\begin{eqnarray}
 \det\left(D_{\rm W}+m_0+i\muq\gamma_5\tau^3\right) &=& 
  \det\left(\begin{array}{cc}
              Q+i\mu_q & 0 \\
              0  & Q-i\mu_q
             \end{array}\right)  \nonumber \\
 &=& \det(Q^2+\mu_q^2) > 0.
\end{eqnarray}
The difference in the determinant already shows that twisted mass and standard QCD
cannot be the same regularisation. In fact, any attempt to perform 
an axial rotation so as to eliminate the twisted mass term will 
rotate the Wilson term in eq.~(\ref{eq:Dw}), too.
the equivalence between both regularisations can therefore only be expected to hold in the 
continuum limit. We will discuss this more in detail below.

Here it suffices to say that the chiral flavour symmetry 
of twisted mass QCD is reduced to an exact U(1) symmetry with generator $\tau^3/2$. 
Furthermore, charge conjugation, axis permutations and reflections
combined with a flavour permutation, e.g.
\begin{equation}
 \psi(x_0,{\bf x}) \rightarrow \gamma_0\tau^1 \psi(x_0,-{\bf x}),\qquad
\psibar(x_0,{\bf x}) \rightarrow \psibar(x_0,-{\bf x})\gamma_0\tau^1,
\end{equation}
are exact symmetries. Finally, the construction of a positive and self-adjoint transfer
matrix for standard Wilson quarks can be generalised to twisted mass QCD, 
provided $\muq$ is real and the usual condition on the standard bare
mass parameter, $|\kappa| < 1/6$, with $\kappa=(2am_0+8)^{-1}$ 
is satisfied~\cite{Frezzotti:2001ea}.

\subsection{Equivalence between tmQCD and QCD}

Taking the continuum limit, we see that the fermionic continuum action of tmQCD,
\begin{equation}
 S_f = \int{\rm d}^4x\,\,
  \psibar(x)(\mbox{$D${\hspace{-2.1mm}/}}+m+i\muq\gamma_5\tau^3)\psi(x),
\end{equation}
can be related to the standard action by a global chiral field rotation,
\begin{equation}
 \psi'=R(\alpha)\psi, \quad \psibar'=\psibar R(\alpha),
  \quad R(\alpha)=\exp\left(i\alpha\gamma_5\frac{\tau^3}{2}\right).
\label{eq:axial}
\end{equation}
Choosing the angle $\alpha$ such that $\tan\alpha = \muq/m$, the action
for the primed fields takes the standard form,
\begin{equation}
 S_f^{\prime} = \int{\rm d}^4x\,\,
  \psibar^{\prime}(x)(\mbox{$D${\hspace{-2.1mm}/}}+M)\psi^{\prime}(x),\qquad
  M = \sqrt{m^2+\muq^2}.
\end{equation}
In QCD all physical observables can be extracted from gauge invariant correlation 
functions of composite fields. We would therefore like to study the relationship between 
correlation functions in tmQCD and standard QCD. To this end we introduce polar mass 
coordinates,
\begin{eqnarray}
 m=M\cos(\alpha),  \qquad  \mu_q=M\sin(\alpha),
\end{eqnarray}
and consider the correlation functions labelled by ($M,\alpha$),
\begin{eqnarray}
 \left\langle O[\psi,\bar\psi] \right\rangle_{(M,\alpha)} 
 = {\mathcal Z}^{-1}\int D[U,\psi,\bar\psi] \  O[\psi,\bar\psi] \ {\rm e}^{-S[m,\muq]}.
\end{eqnarray}
Treating the functional integral like an ordinary integral we change the variables to 
$\psi'$ and $\psibar'$ of eq.~(\ref{eq:axial}) 
and re-label these new integration variables to $\psi$ and $\psibar$
afterwards. In this way we arrive at the identity,
\begin{eqnarray}
 \left\langle O[\psi,\psibar] \right\rangle_{(M,0)} 
 = \left\langle O[R(\alpha)\psi,\psibar R(\alpha)] \right\rangle_{(M,\alpha)}.
 \label{eq:corr_identity}
\end{eqnarray}
To go a step further, we now assume that the functional
$O[\psi,\psibar]$ consists of factors which are members
of a chiral multiplet.
Considering such a field $\phi_A^{(r)}[\psi,\psibar]$ 
in the representation $r$, the transformation of $\psi$ and $\psibar$ by
$R(\alpha)$ induces the transformation of $\phi_A^{(r)}$ by
$R^{(r)}(\alpha)$ in the representation $r$,
\begin{eqnarray}
 \phi_A^{(r)}[R(\alpha)\psi,\psibar R(\alpha)]
 = R_{AB}^{(r)}(\alpha)\phi_B^{(r)}[\psi,\psibar].
\end{eqnarray}
For $n$-point functions of such fields, one obtains the identity,
\begin{eqnarray}
 \langle \phi_{A_1}^{(r_1)}\cdots\phi_{A_n}^{(r_n)}\rangle_{(M,0)}
 =\left(\prod_{i=1}^{n} R_{A_i B_i}^{(r_i)}(\alpha)\right)
   \langle \phi_{B_1}^{(r_1)}\cdots\phi_{B_n}^{(r_n)}\rangle_{(M,\alpha)}.
 \label{eq:relate}
\end{eqnarray}
The correlation functions in standard QCD labelled by $(M,0)$ are just
linear combinations of those in twisted mass QCD, labelled by $(M,\alpha)$.
The inverse relation can be obtained by inverting the 
matrices $R^{(r)}(\alpha)$. This is trivial, as the axial rotation
(\ref{eq:axial}) forms an abelian subgroup of the chiral flavour group,
so that $[R^{(r)}(\alpha)]^{-1}=R^{(r)}(-\alpha)$.
Examples of chiral multiplets are the non-singlet currents $(A_{\mu}^a,V_{\mu}^a)$ 
or the non-singlet axial density  combined with the singlet scalar density, 
$(\frac{1}{2}S^0,P^a)$. In terms of quark fields one has
\begin{eqnarray}
\begin{array}{ll}
 A_{\mu}^a=\psibar\gamma_{\mu}\gamma_5\frac{\tau^3}{2}\psi,
 & \ \ \ \ \ 
 V_{\mu}^a=\psibar\gamma_{\mu}\frac{\tau^3}{2}\psi, \\ 
 P^a=\psibar\gamma_5\frac{\tau^3}{2}\psi,
 & \ \ \ \ \ 
 S^0=\psibar\psi,
\end{array}
\end{eqnarray}
and one may then easily infer the transformation behaviour 
of these chiral multiplets:
\begin{eqnarray}
\begin{array}{ll}
 A_{\mu}^{\prime 1}=cA_{\mu}^{1}+sV_{\mu}^{2},
 & \ \ \ \ \ \ 
 V_{\mu}^{\prime 1}=cV_{\mu}^{1}+sA_{\mu}^{2}, \\
 A_{\mu}^{\prime 2}=cA_{\mu}^{2}-sV_{\mu}^{1},
 & \ \ \ \ \ \ 
 V_{\mu}^{\prime 2}=cV_{\mu}^{2}-sA_{\mu}^{1}, \\
 A_{\mu}^{\prime 3}=A_{\mu}^{3},
 & \ \ \ \ \ \ 
 V_{\mu}^{\prime 3}=V_{\mu}^{3}, \\
 P^{\prime a}=P^{a}, \ \ \ (a=1,2),
 & \ \ \ \ \ \ 
 S^{\prime 0}=cS^0+2isP^3,\\
 P^{\prime 3}=cP^{3}+is\frac{1}{2}S^0.\\
\end{array} \label{eq:map}
\end{eqnarray}
Here the notation $O^{\prime}\equiv O[\psi^{\prime},\bar\psi^{\prime}]$, $c\equiv
\cos(\alpha)$, $s\equiv\sin(\alpha)$ was used.
For a correlator of $A_{\mu}^1(x)$ and $P^1(y)$ in standard QCD this means
\begin{eqnarray}
 \left\langle A_{\mu}^1(x)P^1(y)\right\rangle_{(M,0)} &=& 
 \cos(\alpha)\ \left\langle A_{\mu}^1(x)P^1(y)\right\rangle_{(M,\alpha)} \nonumber \\
 && \mbox{} +\sin(\alpha)\ \left\langle V_{\mu}^2(x)P^1(y)\right\rangle_{(M,\alpha)}.
\end{eqnarray}
In other words, eqs.~(\ref{eq:map}) relate an
insertion of the primed fields into standard QCD correlators
to the insertion of the corresponding r.h.s.~into tmQCD correlators.
In particular, we note that the PCAC and PCVC relations in the physical basis
\begin{equation}
  \partial_{\mu}A_{\mu}^{\prime a} = 2 M P^{\prime a},\qquad
  \partial_{\mu}V_{\mu}^{\prime a} =0,
\end{equation}
are equivalent to linear combinations of their twisted counterparts,
\begin{eqnarray}
 \partial_{\mu}A_{\mu}^a &=& 2mP^a+\delta^{3a}i\mu_q S^0, \nonumber\\
 \partial_{\mu}V_{\mu}^a &=& -2\mu_q\epsilon^{3ab} P^b.
\end{eqnarray}
In conclusion, the formal continuum theory provides us  with a dictionary
between correlation functions in standard and twisted mass QCD. However,
all these considerations have been quite formal, and we need to specify 
how such a dictionary carries over to the renormalized theories.

\subsection{Beyond the formal continuum theory}

To clarify this question let us suppose that tmQCD is regularised 
on the lattice with  Ginsparg-Wilson quarks, where chiral and 
flavour symmetries are the same as in the continuum.
Identities such as Eqs.~(\ref{eq:relate})  may then be 
derived in the {\em bare} theory. If in addition, we start from a finite volume with, 
say, periodic boundary conditions for all fields, 
the functional integral becomes a finite dimensional Grassmann integral.
Therefore, these identities are no longer formal, but on firm mathematical grounds, 
and all one has to show is that the renormalisation procedure 
can be carried out such that they continue to hold in
the renormalised theory. This is straightforward, as one just has to make 
sure that all  members of a given chiral multiplet are renormalised 
in the same way, and that the multiplicative renormalization constants 
do not depend on  the twist angle $\alpha$. This can be achieved e.g.~by imposing 
renormalisation conditions in the massless limit. Hence, in this case, 
the dictionary introduced above holds between the renormalised 
correlation functions of both theories. Assuming universality to hold
beyond perturbation theory, this establishes the equivalence of both
versions of QCD at the non-perturbative level, since any other
regularisation, chirally symmetric or not, will then lead to the 
same renormalised correlation functions up to cutoff effects.
While there is no reason to doubt that universality holds generally, 
one should be aware that it has rigorously been established
only in perturbation theory and for selected regularisations 
(e.g.~lattice regularisations with Wilson type quarks~\cite{Reisz:1988kk}).

\subsection{Lattice tmQCD with Wilson quarks}

In tmQCD on the lattice with Wilson quarks the axial transformation relating
continuum tmQCD to standard QCD is not an exact symmetry.
Therefore, equivalence can only be expected to hold in the
continuum limit, i.e.~for properly renormalized correlation functions
and up to cutoff effects.
The lattice symmetries imply the counterterm structure, with the
following result for the renormalised parameters,
\begin{equation}
 g_{\rm R}^2=Z_g g_0^2, \qquad
 m_{\rm R}=Z_m (m_0-\mc), \qquad
 {\mu}_{\rm R}=Z_{\mu} {\muq}.
\end{equation}
It is a priori not obvious how the twist angle $\alpha$
should be defined from the mass parameters.
The key observation is that chiral symmetry
can be restored in the bare lattice theory up to cutoff effects, by
imposing axial Ward identities as normalisation conditions~\cite{Bochicchio:1985xa}.
This fixes the {\em relative} renormalization of all members of
a chiral multiplet, such as $Z_{\rm A}/Z_{\rm V}$ for the 
symmetry currents\footnote{$Z_{\rm V}=1$ only holds if the 
(partially) conserved point-split vector current $\tilde{V}_\mu^a$ is used.}, 
or $Z_{S^0}/Z_P$ for the iso-triplet axial and the iso-inglet scalar densities.
Note that such ratios are scale independent functions of $g_0$ only,
which are expected to converge to $1$ in the continuum limit 
with a rate $g_0^2 \propto -1/\ln a$. 
In particular, these ratios do not depend upon the quark mass parameters
and may therefore be determined in the massless limit~\cite{Luscher:1996jn} 
where the tmQCD and standard QCD actions coincide.
The connection between the mass parameters and chiral Ward identities 
is established by choosing renormalisation schemes such 
that the PCAC and PCVC relations hold, 
with the renormalised currents and and axial density, and the
renormalised mass parameters. The renormalization constants 
may then be shown to satisfy the identities,
$Z_m=Z_{\rm S^0}^{-1}$ and $Z_\mu=Z_{\rm P}^{-1}$.
With these conventions it is clear that the
ratio of renormalised mass parameters is known once the critical mass and 
the ratio $Z_{S^0}/Z_P$ are given,
\begin{equation}
  \tan\alpha = \frac{\mu_{\rm R}}{m_{\rm R}}
  = \frac{Z_{\rm S^0}}{Z_{\rm P}}\frac{\muq}{m_0-m_{\rm cr}}.
\end{equation}
Besides the ratio of renormalization constants one thus needs to 
determine the critical mass. In practice this can be done by
measuring a bare PCAC mass $m$ from correlation functions with
some external field $O$,
\begin{eqnarray}
 m=\frac{\langle\partial_{\mu}A_{\mu}^1(x)O\rangle}{\langle P^1(x)
  O\rangle},
\end{eqnarray}
and by using the relation 
\begin{equation}
   m_{\rm R}= Z_P^{-1}Z_{\rm A} m,\qquad m=Z_{\rm A}^{-1}Z_m Z_P (m_0-m_{\rm cr}).
\end{equation}
Alternatively one may use the measured bare PCAC quark mass $m$ to
obtain $\alpha$ directly,
\begin{equation}
 \tan\alpha= \muq/(Z_{\rm A} m),
\end{equation}
provided one has previously determined $Z_{\rm A}$. Already at this point
one notes that the choice $\alpha=\pi/2$ is special, as in this case
one merely needs to determine the critical mass. The choice $\alpha=\pi/2$
is referred to a full or maximal twist, because the physcial quark mass
is then entirely defined by the twisted mass parameter $\muq$.

Having determined the twist angle, and the relative renormalizations
within chiral multiplets, chiral symmetry is restored up to cutoff effects
for the correlation functions of members of these multiplets.
In a second step one just needs to make sure that this property of the
bare theory is not compromised by the renormalization procedure, 
i.e.~one is in a similar situation as in the bare theory with
Ginsparg-Wilson quarks. Proceeding in the same way, the formal identities 
of subsect.~4.1  will hold in the renormalised theory.

An important point to notice is that the twist angle $\alpha$ is a 
new parameter which reflects the freedom to choose a direction
in chiral flavour space for the explicit chiral flavour symmetry breaking. 
Our physical interpretation is such that by definition only the
axial generators are broken by the mass term thus defining the residual 
vector symmetry. With Wilson quarks at non-zero $\alpha$ 
there is an additional breaking of flavour symmetry by the Wilson term.
which is expected to disappear in the continuum limit, just like
chiral symmetry is restored with standard Wilson quarks.
In order to define the continuum limit properly one must make sure that
cutoff effects are a smooth function of $\beta=6/g_0^2$.  In general this
can be achieved by taking the continuum limit at constant physical conditions.
For instance one may keep $m_\pi/F_\pi$ constant as $\beta$ is varied. 
However, in tmQCD this observable is a function of two mass parameters,
or, equivalently of one mass parameter and the twist angle. 
It is crucial that the twist angle is kept constant as the continuum limit
is taken, since the twist angle labels different lattice regularisations 
of two-flavour QCD. In particular, if $\alpha$ is changed from 
one $\beta$-value to the next, there is no reason 
to expect a smooth continuum approach and a continuum extrapolation 
may become impossible.

\section{A few applications of tmQCD}

The relations between tmQCD and standard QCD correlation functions
can be used to by-pass certain lattice renormalization problems 
of standard Wilson quarks (cf.~sect.~3).
As the different operators of a continuum chiral multiplet
are not necessarily related by lattice symmetries,
their renormalisation properties can be very different.
Moreover, the renormalisation properties do not change
in the presence of (twisted or non-twisted) mass terms
except when power divergences are present. Excluding
these cases it is thus sufficient to renormalise 
a given composite field in the chiral limit where the actions of tmQCD
and standard Wilson quarks coincide. One may then choose the operator with the
best renormalisation properties that can be related to the 
desired standard QCD operator by the dictionary established earlier.
Moreover, it may not even be necessary to match the operators
directly. In principle, it is enough to match the desired correlation 
function up to cutoff effects. Perhaps these remarks become clearer by going 
through a few examples:

\subsection{Computation of $F_\pi$}

\begin{figure}[ht]
\epsfig{file=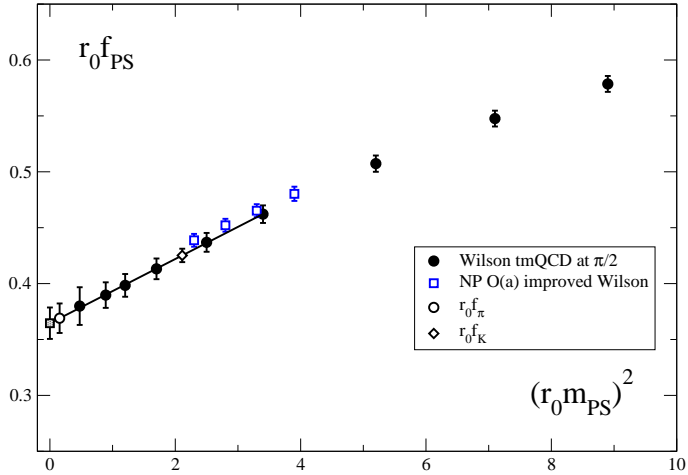,
        clip=,
        angle=270,
        width=11.0cm}%
\caption{Quenched continuum results for $F_\pi$. The plot
also illustrates the absence of the zero mode problem in tmQCD, as much 
smaller pion masses could be reached than with 
standard O($a$) improved Wilson quarks.
\label{fig:fpi}}
\end{figure}

Both the pion mass $m_\pi$ and the pion decay constant $F_\pi$ can
be obtained from the long distance behaviour of the 2-point function
\begin{eqnarray}
 \left\langle(A_{\rm R})_0^1(x)(P_{\rm R})^1(y)a\right\rangle_{(M_{\rm R},0)}
 &=& \cos(\alpha)\
    \left\langle(A_{\rm R})_0^1(x)(P_{\rm R})^1(y)
    \right\rangle_{(M_{\rm R},\alpha)}
 \nonumber \\ &&
 \mbox{}+\sin(\alpha)\
    \left\langle \tilde{V}_0^2(x)(P_{\rm R})^1(y)\right\rangle_{(M_{\rm
    R},\alpha)}.
 \label{eq:2ptfpi}
\end{eqnarray}
The problem with the standard Wilson computation on the l.h.s.~is 
that the axial current requires a non-trivial renormalisation, which
needs to be determined from Ward identities, 
as done e.g.~in~\cite{Luscher:1996jn}.
On the other hand the vector current $\tilde V_\mu^a$ is protected against
such a rescaling since it is conserved at $\muq=0$.
At $\alpha=\pi/2$ the axial current is mapped to the vector current
and one may thus avoid the current renormalisation by computing the
vector correlation function in tmQCD.
It is in fact not necessary to set $\alpha=\pi/2$; when inverting
the relation (\ref{eq:2ptfpi}), 
\begin{eqnarray}
 \left\langle(\tilde{V}_0^2(x)(P_{\rm R})^1(y)\right\rangle_{(M_{\rm R},\alpha)}
 &=& \cos(\alpha)\
    \left\langle(\tilde{V}_0^2(x)(P_{\rm R})^1(y)\right\rangle_{(M_{\rm R},0)}
 \nonumber \\ &&
 \mbox{}+\sin(\alpha)\
    \left\langle(A_{\rm R})_0^1(x)(P_{\rm R})^1(y)\right\rangle_{(M_{\rm
    R},0)}.
\end{eqnarray}
one notices that the first term on the r.h.s~violates both parity and
flavour symmetry of standard QCD. On the lattice this correlation
function therefore contributes at most an O($a$) effect. 
One may thus obtain $F_\pi$ at values $\alpha\ne\pi/2$ 
by computing the l.h.s~of this equation.
Finally, it should be mentioned 
that the exact PCVC lattice relation,
\begin{equation}
 \partial_\mu^\ast \tilde V_\mu^a=-2\muq \varepsilon^{3ab} P^b,
\end{equation}
may be used to replace the vector current by the axial density.
Summing over ${\bf x}$, translation invariance eliminates the spatial
part of the divergence, and the time derivative 
reduces to a multiplication by $m_\pi$ at large time 
separations~\cite{Frezzotti:2001du}.
The results of a quenched computation along these 
lines~\cite{Jansen:2003ir,Jansen:2005kk} are
shown in figures \ref{fig:fpi} and \ref{fig:fpi_scaling}.

\subsection{Direct determination of the chiral condensate}

A computation of the chiral condensate from
the local scalar density has never been performed with Wilson quarks,
due to the cubic divergence (\ref{eq:cubic}) which persists in the chiral limit.
In tmQCD the r\^ole of the scalar density is played by the 
axial density, i.e.~one expects the relation
\begin{equation}
  \left\langle (P_{\rm R})^3(x)\right\rangle_{(M_{\rm R}, \alpha)} = 
  \cos(\alpha) \left\langle (P_{\rm R})^3(x)\right\rangle_{(M_{\rm R}, 0)} 
  -\frac{i}{2}\sin(\alpha) 
  \left\langle (S_{\rm R})^0(x)\right\rangle_{(M_{\rm R}, 0)} 
\end{equation}
Again, the first term on the r.h.s.~vanishes up to O($a$) due to parity,
so that the computation of the l.h.s.~yields the chiral condensate up 
to the factor $(-i/2)\sin(\alpha)$. This is advantageous as the renormalised
axial density is of the form,
\begin{equation}
  (P_{\rm R})^3 = Z_{\rm P}\left(P^3 + \muq c_{\rm P}\ a^{-2}\right),
\end{equation}
i.e.~the power divergence vanishes for $\muq=0$.
Still, in order to determine the condensate one needs to perform 
first the infinite volume limit followed by the $\muq=0,m_0=m_{\rm cr}$ limits 
(at fixed $\alpha$) and the continuum limit, which remains a rather delicate task. 
In particular, the chiral limit is complicated by
the fact that the uncertainty in $\sin(\alpha)$ increases as the 
quark mass is decreased, due to the intrinsic O($a$) ambiguity of $\mc$. 
In practice this means that one has to extrapolate to the 
chiral limit from some distance, but this is anyway required 
for finite volume effects to remain small.

\subsection{The computation of $B_{K}$}

Four-quark operators provide an interesting playground for 
mappings between tmQCD and standard QCD.
We start with the $B_K$ parameter which is defined in QCD with 
dynamical $u, d, s$ quarks by,
\begin{eqnarray}
 \langle\bar{K}^0\mid O_{(\rm V-A)(\rm V-A)}^{\Delta S=2}\mid
  K^0\rangle 
 = {\textstyle \frac{8}{3}}F_{K}^2 m_{K}^2 B_{K}.
\end{eqnarray}
The local operator
\begin{eqnarray}
 O_{(\rm V-A)(\rm V-A)}^{\Delta S=2}
 =\sum_{\mu}[\bar{s}\gamma_{\mu}(1-\gamma_5)d]^2,
\end{eqnarray}
is the effective local interaction induced by integrating out the
massive gauge bosons and $t$-,$b$- and $c$-quarks in the Standard Model.
The transition between the pseudoscalar states $K^0$ and $\bar{K}^0$ 
does not change parity. Therefore, only the parity-even part in the effective
operator,
\begin{eqnarray}
 O_{(\rm V-A)(\rm V-A)}
 =\underbrace{O_{\rm VV+AA}}_{\mbox{parity-even}}
   -\underbrace{O_{\rm VA+AV}}_{\mbox{parity-odd}},
\end{eqnarray}
contributes to $B_K$. With Wilson type quarks, the operators $O_{\rm VV+AA}$ and
$O_{\rm VA+AV}$ are renormalised as follows
\begin{eqnarray}
 (O_{\rm VV+AA})_{\rm R} &=&
 Z_{\rm VV+AA}\Bigl\{O_{\rm VV+AA}+\sum_{i=1}^{4}z_i \ O_i^{d=6}\Bigr\},
 \\
 (O_{\rm VA+AV})_{\rm R} 
&=& Z_{\rm VA+AV}O_{\rm VA+AV}.
\end{eqnarray}
While the parity-even component mixes with four other 
operators of dimension 6, the parity-odd
component only requires multiplicative renormalisation, due to CP and
flavour exchange symmetries~\cite{Bernard:1987pr}. 
This raises the question if one can by-pass the
mixing problem by exchanging the r\^oles of both operators
through the introduction of twisted mass terms. 
This is indeed possible, but one first needs to introduce
the strange quark. The simplest possibility consists in adding 
a standard $s$-quark to a twisted quark doublet $\psi$ of the light 
up and down quarks, which are thus taken to be degenerate.
The corresponding continuum Lagrangian is given by
\begin{eqnarray}
  {\mathcal L}= \bar\psi(\mbox{$D${\hspace{-2.1mm}/}}+m+i\mu_q\gamma_5\tau^3)\psi
  + \bar{s}(\mbox{$D${\hspace{-2.1mm}/}}+m_s)s,
\end{eqnarray}
and, passing to the physical basis of primed fields, 
one finds
\begin{eqnarray}
 O_{\rm VV+AA}^{\prime} 
 &=& \cos(\alpha)O_{\rm VV+AA}-i\sin(\alpha)O_{\rm VA+AV} \nonumber\\
 &=& -iO_{\rm VA+AV} \ \ \ \ (\alpha=\pi/2).
\end{eqnarray}
At full twist, we thus get a direct mapping between both operators, 
i.e.~$O_{\rm VA+AV}$ in twisted mass QCD at $\pi/2$ is interpreted as 
$O_{\rm VV+AA}$ in standard QCD.  A second possibility consists in
exchanging the r\^oles of up and strange quark, i.e.~one 
considers a twisted doublet of strange and down quarks and a
standard $u$-quark. In this case one finds
\begin{eqnarray}
 O_{\rm VV+AA}^{\prime} 
 &=& \cos(2\alpha)O_{\rm VV+AA}-i\sin(2\alpha)O_{\rm VA+AV} \nonumber\\
 &=& -iO_{\rm VA+AV} \ \ \ \ (\alpha=\pi/4),
\end{eqnarray}
i.e.~the same mapping is obtained, but with the twist angle $\alpha=\pi/4$.
Several comments are in order: while both options, referred
to as $\pi/2$ and $\pi/4$ scenarios respectively, are possible,
the second one is clearly more remote from reality, as it assumes
mass degenerate down and strange quarks.
However, this is precisely the limit in which most lattice calculations
to date have been performed. The justification rests on chiral perturbation theory 
where a a weak dependence upon the strange-down mass difference is predicted.
Moreover, in the quenched approximation, any deviation from the degenerate case
leads to an unphysical logarithmic quark mass dependence~\cite{Sharpe:1994dc,Golterman:1997wb}.

\subsubsection{Renormalisation of $O_{\rm VA+AV}$}

\begin{figure}[ht]
\epsfxsize=10cm   
\epsfbox{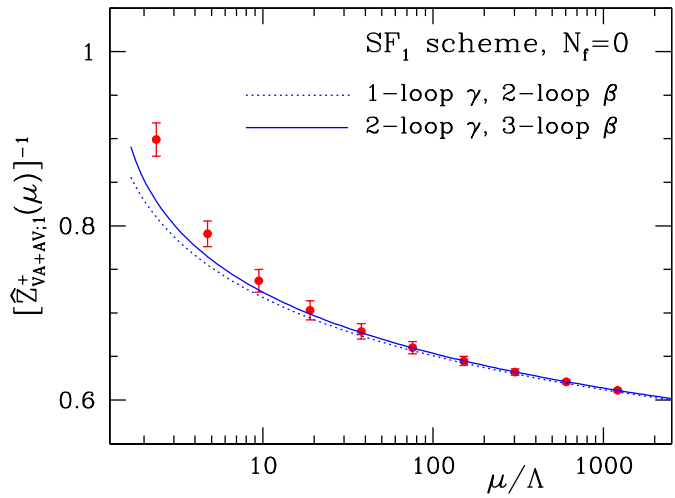}
\caption{The data points show the non-perturbatively 
computed scale evolution of $B_K$ in the
SF scheme. Also shown are two perturbative approximations. \label{fig:runningBK}}
\end{figure}

Whatever the chosen strategy, the operator which requires renormalisation
is $O_{\rm VA+AV}$. The renormalisation is multiplicative, and
the general strategy of \cite{Jansen:1995ck} can be applied.
The scale evolution of the operator in a few Schr\"odinger functional schemes
has been traced in the quenched approximation over a wide range of scales
(for first results with $\Nf=2$ sea quarks cf.~\cite{Dimopoulos:2006es}).
The result is shown in figure \ref{fig:runningBK}. 
It thus remains to calculate the bare matrix element
for $B_K$ at various values of $\beta$, and, after multiplication
with the $Z$-factor at the low energy scale, perform the continuum
limit extrapolation. In the continuum limit one may then use the
known scale evolution to reach the truly perturbative regime
where contact is made with the perturbative renormalisation schemes
of the continuum.
\subsubsection{Results for $B_K$ in the quenched approximation}

Both scenarios have been implemented
in the quenched approximation with lattice spacings 
$a=0.05-0.1$ ${\rm fm}$ and lattice sizes up to $L/a=32$~\cite{Dimopoulos:2006dm}. 
If one sticks to mass degenerate down and strange quarks the $\pi/2$ 
scenario requires some chiral extrapolation, 
due to  the problem with unphysical zero modes (recall that 
the $s$-quark remains untwisted). In the $\pi/4$ scenario the
zero mode problem is eliminated and the kaon mass can be reached
by interpolation, provided the finite volume effects are 
small enough. This is the case with all lattice spacings except
the finest one, where some extrapolation is required.
A combined continuum extrapolation to both data sets, linear in $a$,
leaving out the data at the coarsest lattice spacing led to the 
result $\hat{B}_K=0.789(46)$~\cite{Dimopoulos:2006dm},
where $\hat{B}_K$ denotes the renormalisation group invariant $B$-parameter.
Unfortunately, the twist angle at $\beta=6.1$ had not been tuned
precisely enough, a fact that was only noticed after publication 
of~\cite{Dimopoulos:2006dm}. A new analysis indicates that higher 
than linear lattice artefacts are still significant at $\beta=6.1$. 
As the data set is not sufficient to fit to both $a$ and $a^2$ terms, 
it was decided to discard the data at $\beta=6.1$, too, 
with the result~\cite{Pena:2006tw} 
\begin{equation}
 \hat{B}_K=0.735(71) \qquad \Leftrightarrow\qquad 
 B_K^{\overline{\rm MS}}(2\,{\rm GeV})=0.534(52),
\end{equation}
which is compatible with the earlier result, albeit with a larger uncertainty.
\begin{figure}[ht]
\epsfxsize=10cm   
\epsfbox{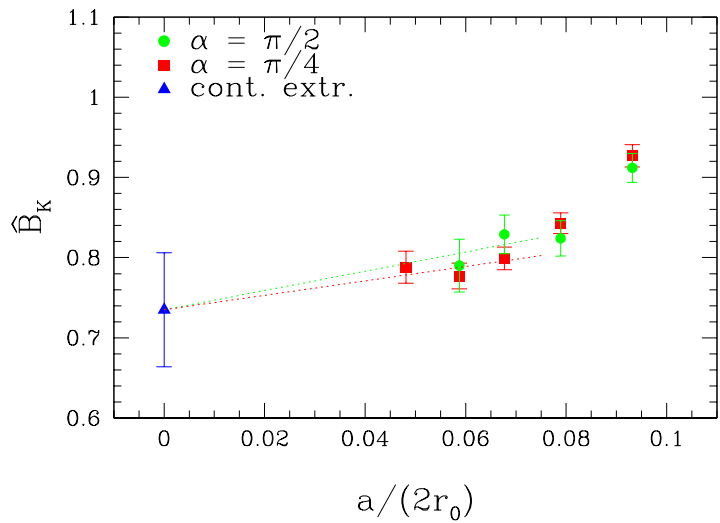}
\caption{Quenched lattice data for both scenarios, $\alpha=\pi/2$ and $\alpha=\pi/4$.
Also shown is the combined continuum extrapolation, leaving out the data 
at the two coarsest lattice spacings.\label{fig:bk}}
\end{figure}
In conclusion, the quenched result for $B_K$ has a total error
of almost 10 percent, which includes all systematic effects
(renormalisation, chiral inter- or extrapolations, continuum extrapolation)
except quenching and the fact that the valence quarks are 
mass degenerate. However a variation of the mass difference up to 
$(M_s-M_d)/(M_s+M_d)\approx 0.5$ did not show sizeable effects.
While the error could still be improved by including data at a 
finer lattice spacing it seems fair to say that further 
progress requires the inclusion of sea quark effects.

\subsection{Further applications}

Twisted mass QCD does not provide a general recipe for
by-passing the lattice specific renormalisation problems of Wilson quarks.
Rather, one needs to discuss on a case by case basis whether it can be 
advantageous to use some variant of tmQCD. 
For further applications to four-quark operators 
and $K\rightarrow \pi$ transitions
I refer the reader to~\cite{Pena:2004gb,Frezzotti:2004wz}.
While the first reference insists on an equal treatment
of sea and valence quarks, the second paper explores a 
mixed action approach, where the valence quarks are chirally
twisted individually, independently of the sea quark action.
This yields a  much greater flexibility and allows for 
a complete elimination of lattice specific mixings and subtractions,
even including O($a$) improvement.
Finally, similar considerations apply
to QCD with static $b$-quarks, where the mixing of four-quark
operators is considerably simplified by twisting the light quarks 
(see~\cite{Pena:2006tw} for a recent review and further references).

\section{$O(a)$ improvement and tmQCD}

Given that the quenched approximation is currently
being overcome, and the zero mode problem for algorithms
can be alleviated, there remain essentially two arguments
in favour of tmQCD as opposed to standard or O($a$) improved Wilson quarks:
the first consists in the possibility to by-pass renormalisation
problems, as explained in the preceding section. The second,
is the property of ``automatic O($a$) improvement" at maximal
twist (i.e.~$\alpha=\pi/2$), as first observed by Frezzotti and 
Rossi~\cite{Frezzotti:2003ni}.
I will explain this point more in detail below, 
after a brief reminder of the situation with standard Wilson quarks.

\subsection{O($a$) improvement of Wilson quarks}

In lattice QCD with Wilson quarks, results are typically 
affected by O($a$) lattice effects, which is to be contrasted 
with staggered or Ginsparg-Wilson quarks where the leading cutoff 
effects are quadratic in $a$.
As illustrated by the $B_K$ determination described above, 
linear lattice artefacts render continuum 
extrapolations more difficult, and it would be nice to get rid of 
them altogether. This is possible by introducing O($a$) counterterms to the action
and the composite operators such that  O($a$) effects are cancelled 
in on-shell quantities. The basic idea goes back to Symanzik~\cite{Symanzik:1983dc},
while the restriction to on-shell quantities in gauge theories has been first
advocated by L\"uscher and Weisz~\cite{Luscher:1984xn}.
When applied to Wilson quarks~\cite{Sheikholeslami:1985ij}, it turns out that
O($a$) improvement of the spectrum (particle masses
and energies) can be achieved by adding a single
counterterm to the action, the so-called Sheikholeslami-Wohlert (SW)
or clover term,
$  i\psibar \sigma_{\mu\nu} F_{\mu\nu} \psi $,
where $F_{\mu\nu}$ is the gluon field tensor
\footnote{On the lattice the field tensor is usually discretised 
using four plaquette terms in the $(\mu,\nu)$-plane whence
the name ``clover term''.}. This term is of dimension five 
and therefore comes with an explicit factor $a$ when included in the
lattice action density. 

While O($a$) improvement of the spectral quantities is 
quite economical, one is often  interested in matrix elements of 
composite operators, and each operator comes with its own set
of O($a$) counterterms, all of which have to be tuned in order
to cancel the linear lattice artefacts. While this may still 
be possible for quark bilinear operators, the counterterms
quickly proliferate in the case of 4-quark operators, and
O($a$) improvement becomes completely impractical if
the quarks are taken  to be mass non-degenerate~(cf.~\cite{Bhattacharya:2005rb}).

\subsection{Automatic O($a$) improvement of tmQCD in a finite volume}

The Symanzik effective theory can also be applied to tmQCD
and a list of O($a$) counterterms for the action and a few
quark bilinear operators can be found in~\cite{Frezzotti:2001ea}. 
The observation in~\cite{Frezzotti:2003ni} is, that at maximal twist,
all the O($a$) counterterms become irrelevant in the sense that
they can at most contribute at O($a^2$). The argument 
for automatic O($a$) improvement can be made such that
it only relies on Symanzik's effective continuum 
theory~\cite{ego,Shindler:2005vj,Frezzotti:2005gi,Aoki:2006nv}
To simplify the discussion, let us first assume that the space-time volume is
finite, so that spontaneous symmetry breaking is excluded 
and all observables are analytic in the quark mass parameters. 
We furthermore assume that we have tuned some PCAC current quark mass
$m_{\rm PCAC}=0$, i.e.~the renormalized standard mass parameter vanishes up
to $O(a)$ effects. Then Symanzik's effective continuum action is given by
\begin{eqnarray}
 S_{\rm eff}=S_0+aS_1+O(a^2), \qquad
 S_0=\int{\rm d}^4x\ \psibar
  (\mbox{$D${\hspace{-2.1mm}/}}+i\mu_q\gamma_5\tau^3)\psi,
\end{eqnarray}
where $S_0$ is the maximally twisted tmQCD continuum action.
$S_1$ is given by 
\begin{eqnarray}
 S_1=\int{\rm d}x^4\ \{c\ i\psibar\sigma_{\mu\nu}F_{\mu\nu}\psi
                       +b_{\mu}\mu^2\psibar\psi +\ldots\},
\end{eqnarray}
where the dots stand for further operators of dimension 5 (possibly
including explicit mass factors), which share the
symmetries of the lattice action. The reason why I omitted them 
here is that they can be eliminated by the equations of motion. Furthermore, 
the second operator can be absorbed in an O($a$) shift of
the standard quark mass parameter, so that one is really left with
the SW term as the only relevant operator for on-shell improvement.
Renormalised (connected) lattice correlation functions can be analysed 
in the effective theory,
\begin{eqnarray}
 \langle O\rangle=\langle O\rangle^{\rm cont}-a\langle S_1 O\rangle^{\rm cont}
                  +a\langle \delta O\rangle^{\rm cont} +O(a^2),
\end{eqnarray}
where the cutoff dependence is explicit. We are here only interested
in the leading cutoff effects at O($a$). To this order there are two contributions, 
first the insertion of the O($a$) part of the effective action $S_1$, 
and second the field specific counterterms $\delta O$. 
For example, with the choice
\begin{eqnarray}
 O=V_{\mu}^1(x)P^2(y),
\end{eqnarray}
one finds the counterterm,
\begin{eqnarray}
 \delta O = \{c_{\rm V}i\partial_{\nu}T_{\mu\nu}^1(x)+\tilde{b}_{\rm
  V}\muq A_{\mu}^2(x)\}P^2(y) + \ldots,
\end{eqnarray}
where the dots stand for further terms which vanish by the equations of motion 
It should be emphasised that the O($a$) (and higher) corrections in the effective
action are only treated as insertions, i.e.~the expectation 
values $\langle\cdot\rangle^{\rm cont}$ are taken with respect 
to the continuum action $S_0$. In writing down the 
effective Symanzik theory there is thus an implicit 
assumption made, namely that one is working in 
the regime of continuum QCD where cutoff effects 
only appear as asymptotically small corrections.
This assumption may certainly be wrong in some regions of 
parameter space, and particular care has to be taken in the presence 
of phase transitions.

To proceed I introduce the $\gamma_5\tau^1$-transformation,
\begin{equation}
 \psi\rightarrow i\gamma_5\tau^1\psi, \qquad 
  \psibar\rightarrow \psibar i\gamma_5\tau^1,
 \label{eq:gam5tau1}
\end{equation}
which is part of the vector symmetry
of two-flavour QCD. Hence $S_0$ is invariant,
but this is not the case for $S_1$, i.e.~one finds
\begin{equation}
 S_0 \rightarrow S_0, \qquad S_1 \rightarrow -S_1.
\end{equation}
For gauge invariant fields the transformation (\ref{eq:gam5tau1}) squares
to the identity, so that one may define an associated parity.
For fields $O$ with a definite $\gamma_5\tau^1$-parity one then finds,
\begin{equation}
 O \rightarrow \pm O  \quad\Rightarrow\quad
 \delta O \rightarrow \mp \delta O.
\end{equation}
By applying the $\gamma_5\tau^1$ transformation to
the integration variables in the functional integral, one
may derive identities between correlation functions, due
to the invariance of the continuum action and functional measure.
In particular, if we choose a $\gamma_5\tau^1$-even field $O$,
we find for the correlation functions at O($a$) 
\begin{eqnarray}
 \langle S_1 O\rangle^{\rm cont} &=& -\langle S_1 O\rangle^{\rm cont} =0, \nonumber\\
 \langle \delta O\rangle^{\rm cont} &=& -\langle \delta O\rangle^{\rm cont}=0,
\end{eqnarray}
and therefore
\begin{eqnarray}
 \langle O\rangle &=& \langle O\rangle^{\rm cont} +O(a^2).
\end{eqnarray}
For a $\gamma_5\tau^1$-odd $O$, one obtains
\begin{eqnarray}
 \langle O\rangle^{\rm cont} &=& -\langle O\rangle^{\rm cont}=0, \nonumber\\
 \langle S_1 O\rangle^{\rm cont} &=& \langle S_1 O\rangle^{\rm cont}, \nonumber\\
 \langle \delta O\rangle^{\rm cont} &=& \langle \delta O\rangle^{\rm cont},
\end{eqnarray}
which implies
\begin{equation}
 \langle O\rangle = -a\langle S_1 O\rangle^{\rm cont}
                      +a\langle \delta O\rangle^{\rm cont} +O(a^2).
\end{equation}
We may thus conclude that, at least in a small finite volume
lattice correlation functions of $\gamma_5\tau^1$-even fields are 
automatically $O(a)$ improved, while those of $\gamma_5\tau^1$-odd 
fields vanish up to $O(a)$ terms. As a corollary, one may state
that standard Wilson quarks in a finite volume are automatically O($a$) improved
in the chiral limit. Although this is not the most interesting regime of QCD,
it is somewhat surprising that this fact had not been noticed for
more than 2 decades! To conclude this section note 
that in terms of the physical basis, (\ref{eq:gam5tau1}) corresponds to
the discrete flavour transformation,
\begin{equation}
 \psi^{\prime}\rightarrow -i\tau^2\psi^{\prime}, \qquad
 \psibar^{\prime}\rightarrow \psibar^{\prime}i\tau^2.
\end{equation}
A very similar argument based on parity transformations 
has been  given by Shindler in~\cite{Shindler:2005vj}.
In \cite{Aoki:2006nv} a systematic analysis of the $\gamma_5\tau^1$ symmetry
(called $T_1$ in this paper) can be found, 
showing that not only O($a$) but all odd powers of $a$ vanish
in $\gamma_5\tau^1$-even correlators. This is not surprising, as this is
implicit in the earlier analysis in \cite{Frezzotti:2003ni}, where the same conclusion was drawn.

\subsubsection{Uncertainty of the chiral limit}

If O($a$) improvement is automatic one might think that it should be possible
to determine the critical mass $m_{\rm cr}$ up to an intrinsic O($a^2$) uncertainty. 
This is not so, as I will now explain. The critical mass can be determined
by tuning some PCAC mass to zero, and there is no obstacle for doing
this in a finite volume.  Now, the PCAC relation involves 
the axial current and density, $A_{\mu}^a$ and $P^a$, 
which have opposite $\gamma_5\tau^1$-parities. 
According to the preceding discussion this means, for the first flavour components
and with a $\gamma_5\tau^1$-even source field $O_{\rm even}$,
\begin{eqnarray}
 \langle \partial_{\mu}A_{\mu}^1(x)O_{\rm even}\rangle =
 2\underbrace{m_{\rm PCAC}}_{\mbox{O($a$)}}
 \underbrace{\langle P^1(x)O_{\rm even}\rangle}_{\mbox{O($a$)}} = O(a^2).
\end{eqnarray}
The l.h.s. being $\gamma_5\tau^1$-even must vanish up to O($a^2$), provided
maximal twist is realised at least up to cutoff effects, i.e.~$m_{\rm R}={\rm O}(a)$.
This implies that the PCAC mass is of O($a$), too, multiplying
a correlation function which is $\gamma_5\tau^1$-odd and therefore of O($a$).
Thus no contradiction arises, the O($a^2$) of the l.h.s.~is matched
on the r.h.s.~by two factors of O($a$).

Another way to understand that an O($a$) shift in the critical mass
does not ruin O($a$) improvement is to treat such a shift as
an insertion of the standard mass operator $\psibar\psi$ into
correlation functions. This operator is $\gamma_5\tau^1$-odd
so that its insertion into a $\gamma_5\tau^1$-even correlator
produces an O($a$) effect, which together with the 
O($a$) mass shift yields an O($a^2$) effect.

\subsection{Automatic O($a$) improvement in infinite volume}

\begin{figure}[ht]
\epsfig{file=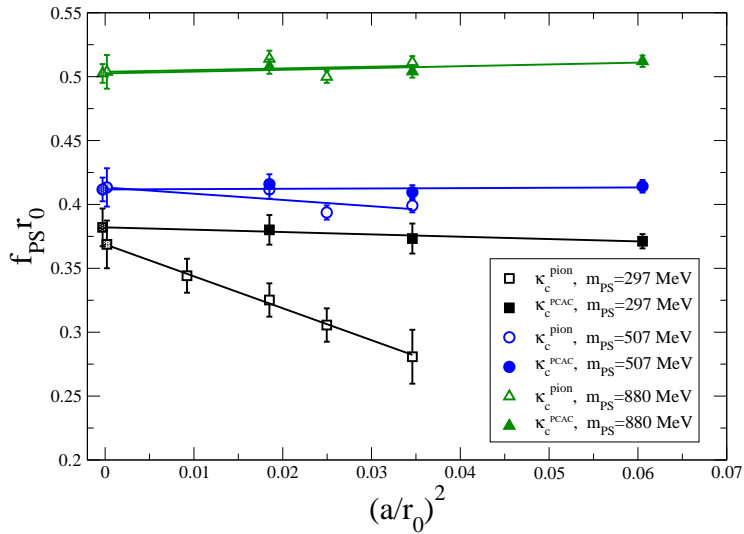,
        clip=,
        angle=270,
        width=11.0cm}%
\caption{The continuum approach of $F_\pi$ in quenched tmQCD for 
various pion masses vs.~$a^2/r_0^2$.  \label{fig:fpi_scaling}}
\end{figure}

When the infinite volume limit is taken, the basic difference
is the presence of spontaneous symmetry breaking and the appearance of
non-analyticities in the mass parameters near the chiral limit. 
As discussed earlier, twisted mass QCD is a valid regularisation of 
two-flavour QCD provided the continuum limit is taken at fixed twist angle. 
To maintain maximal twist, i.e.~$\alpha=\pi/2$ one needs 
to tune the standard quark mass to $m_{\rm cr}$, which has an intrinsic
O($a$) ambiguity. As long as the twisted mass is much larger
than the typical O($a$) spread of $\mc$, the twist angle may 
be considered well-defined, and the continuum limit is reached 
with O($a^2$) corrections.
However, in practice one is interested in varying the quark mass at
fixed cutoff, rather than studying the quark mass dependence only
in the continuum limit. Approaching the chiral limit at fixed $a$ by lowering
the twisted mass one enters the regime where the twisted mass parameter
becomes comparable to the O($a$) ambiguity of $\mc$.
One may debate at this point whether the relevant comparison is with
the uncertainty of $\mc$ itself or rather with the size of typical 
O($a^2$) effects in correlation functions generated by this uncertainty.
In any case one reaches a point where the control over the twist angle
is lost. When delivering my Nara lectures I interpreted this fact
as a breakdown of the effective Symanzik theory. This is perhaps too
rigid an interpretation. Rather one could say that 
for every definition of $\mc$, an effective twist angle is formed  
by the dynamics of the system, which may be far from the maximal twist 
one would like to maintain. Moreover, without further input is is impossible 
to know the effective twist angle for a given definition of $\mc$.
This is a disaster, as the whole interpretation of the theory rests 
on the twist angle, and a change in the effective twist angle 
(which remains unnoticed!) might strongly affect some correlators even at O(1)! 
Fortunately this problem occurs close to the chiral limit, and thus in a region 
of parameter space where Chiral Perturbation Theory ($\chi$PT) is expected
to describe the dynamics in terms of pion physics \cite{Sharpe:2006pu}. 
In particular, $\chi$PT is able to identify definitions of 
$\mc$ in terms of pionic observables, which lead to an
effective twist angle of $\alpha=\pi/2$, so that the Symanzik 
effective theory for maximally twisted mass QCD remains applicable
in this region. For instance, this should be the case 
if one requires parity or flavour symmetry restauration, 
e.g.~by imposing that a $\gamma_5\tau^1$-odd pion correlation 
function vanishes. Note that the vanishing of
the PCAC mass for a pion correlation function is a special case
of such a condition. On the other hand, according to \cite{Aoki:2004ta,Aoki:2006nv}
the condition of vanishing pion mass (calculated in the untwisted theory)
does indeed lead to a O(1) variation of the effective twist angle.
However, apart from larger cutoff effects of O($a^2$) this does
not (yet?) seem to be a major problem in~\cite{Jansen:2005kk}, 
cf.~figure \ref{fig:fpi_scaling}.
In any case, as the spontaneous symmetry breaking is closely 
related to the dynamics of pions, it seems that no statement can be made about
generic definitions of $\mc$ in a small volume, 
either~from axial current conservation, or from parity or 
flavour symmetry restauration.

\section{Consequences of Parity and Flavour breaking}

The exact symmetries of lattice QCD with standard
Wilson quarks include parity and flavour symmetry 
which are used to classify the hadron spectrum.
This is very convenient in any hadron analysis: even at
fixed lattice spacing, the excited states which may occur in a given channel
can be read from the Particle Data Book, with the exception 
of states with higher spin and/or angular momentum 
where the correspondence is spoilt by the lack of rotational symmetry on 
the lattice.

The situation is different in tmQCD since both parity and flavour symmetry 
are broken by the Wilson term. As a consequence, the classification
in isospin multiplets fails by terms of O($a$), or O($a^2$) if
O($a$) improvement is at work. For instance, the neutral pion
is not mass degenerate with the charged pions, or 
the nucleon $\Delta$-resonances no longer form an exact isospin multiplet.
Various simulations of quenched tmQCD have confirmed these expectations,
and point to a restoration of flavour symmetry in
the continuum limit~\cite{Abdel-Rehim:2005gz,Abdel-Rehim:2006ve,Jansen:2005cg}, 
although the expected  rate $\propto a^2$ for maximally twisted mass QCD
has not in all cases been demonstrated convincingly.

However, the splittings of isospin multiplets by cutoff effects
are not the most serious drawback of parity and flavour symmetry breaking.
In the spectral analysis of a hadronic two-point function all
excited states with the same lattice quantum numbers may contribute.
Even though the states violating continuum symmetries are
multiplied by coefficients proportional to $a$, these
states need to be taken into account when working at fixed lattice
spacing. Particularly annoying is the neutral pion, which shares
all the lattice quantum numbers with the vacuum. One may thus
add a neutral pion to any state without changing its lattice
quantum numbers. The presence of additional relatively
light states may require a multistate analysis just to identify
and subtract states which are a pure lattice artefact. Moreover,
correlation functions involving the light pion require the evaluation
of disconnected diagrams. However, it should be emphasised that 
these problems are purely technical; conceptually
tmQCD is on a very solid basis, and in contrast to staggered 
fermions there is no mixing between flavour and spin degrees of freedom.

\subsection{Non-degenerate quarks and additional flavours}

Twisted mass QCD was originally formulated for a single doublet of 
mass degenerate flavours. This can easily be generalised 
to include more mass degenerate doublets. However,
such a spectrum is quite unrealistic unless 
a non-degeneracy can be introduced within a doublet.
Moreover, this non-degeneracy should not cause 
too much damage to all the nice properties
of tmQCD. In particular, one needs to maintain the reality and
positivity of the quark determinant, if such an action is
to be used for simulations of full tmQCD.
This is indeed possible, by introducing
a mass splitting term as follows~\cite{Frezzotti:2003xj},
\begin{eqnarray}
 {\mathcal L}=  \psibar(\mbox{$D${\hspace{-2.1mm}/}}+m+i\muq\gamma_5\tau^3
    +\delta_m\tau^1)\psi,
\end{eqnarray}
where $\delta_m$ is the mass splitting parameter. 
The mass spectrum is easily obtained by going to the 
physical basis and diagonalising the mass matrix.
Its eigenvalues are then found to be $M_{\pm}=\sqrt{m^2+\muq^2} \pm \delta_m$.
Translating this continuum situation to Wilson quarks in
the obvious way, one first notices that the determinant 
of the twisted Wilson-Dirac operator must be real due to 
the conjugation property,
\begin{eqnarray}
 && \gamma_5\tau^1\left
   (D_{\rm W}+m_0+i\mu_q\gamma_5\tau^3+\delta_m\tau^1\right)\gamma_5\tau^1 \nonumber\\
 && \hphantom{0123}=\left(D_{\rm W}+m_0+i\mu_q\gamma_5\tau^3+\delta_m\tau^1\right)^\dagger.
\end{eqnarray}
Furthermore, the flavour structure of the determinant
can again be reduced analytically, with the result, 
\begin{eqnarray} 
 && \det(D_{\rm W}+m_0+i\mu_q\gamma_5\tau^3+\delta_m\tau^1) \nonumber \\
 &&\hphantom{0123}= \det(Q^2+ \delta_m[\gamma_5,Q] +\mu_q^2-\delta_m^2),
\end{eqnarray}
and this determinant is non-zero provided $\muq^2 >\delta_m^2$. 
The positivity of the determinant at $\delta_m=0$ and continuity
in $\delta_m$ then imply positivity of this determinant for  
non-zero $\delta_m$.

The mass splitting parameter is renormalised multiplicatively
$ \delta_{m,{\rm R}}=Z_{\rm S}^{-1}\delta_m $, 
where $Z_{\rm S}$ is the renormalisation constant of the non-singlet
scalar density. As the positivity of the determinant follows
from a condition on the bare parameters $\muq$ and $\delta_m$, 
the corresponding condition in terms of the renormalised parameters
involves a ratio of renormalisation constants, i.e. 
$ \delta_{m,{\rm R}} < (Z_{\rm P}/Z_{\rm S})\mu_{\rm R}$.
The value of $Z_{\rm P}/Z_{\rm S}$ depends on details of the
regularization, so that one cannot make a general statement
about the ensuing limitations (if any).
However, it is remarkable that one may use this action
to perform numerical simulations with two non-degenerate light
quark flavours, as needed for instance to study small isospin 
breaking effects.
If used for strange and charm quarks~\cite{Chiarappa:2006ae,Jansen:2006rf}, however, 
one potentially has to deal with a fine tuning problem for the strange quark mass:
for instance, assuming $m_s=100$ ${\rm  MeV}$ and $m_c=1300$ ${\rm MeV}$, these
values are obtained as $(700\pm 600)$ ${\rm MeV}$.
Finally, it should be said that the presence of the additional flavour
non-diagonal breaking term renders the relationship to standard QCD more 
complicated, and the flavour structure needs to be dealt with explicitly 
in numerical calculations of quark propagators.

\section{A chiral twist to the QCD Schr\"{o}dinger functional} 

In order to solve scale dependent renormalisation problems
the introduction of an intermediate renormalisation scheme
based on the Schr\"odinger functional (SF scheme) is an attractive
possibility\cite{Jansen:1995ck}. Here I start by summarising its basic features
in order to prepare the discussion of possible improvements.

\subsection{The QCD Schr\"{o}dinger functional}

The QCD Schr\"odinger functional\cite{Luscher:1992an,Sint:1993un}(SF) 
is the functional integral for QCD where the Euclidean space-time manifold 
is taken to be a hyper cylinder.
The quantum fields are periodic in space, and Dirichlet conditions
are imposed at (Euclidean) times $x_0=0$ and $x_0=T$.
\begin{eqnarray}
 P_{+}\psi(x)\mid_{x_0=0}&=&\rho, \ \ \ \ \ 
 P_{-}\psi(x)\mid_{x_0=T}=\rho^{\prime}, \nonumber\\
 \bar\psi(x)P_{-}\mid_{x_0=0}&=&\bar\rho, \ \ \ \ \ 
 \bar\psi(x)P_{+}\mid_{x_0=T}=\bar\rho^{\prime}, \nonumber\\
 A_k(x)\mid_{x_0=0}&=&C_k, \ \ \ \ \ 
 A_k(x)\mid_{x_0=T}=C_k^{\prime}, \quad k=1,2,3,
\end{eqnarray}
with the projectors $P_\pm=\frac12(1\pm\gamma_0)$.
Correlation functions are then defined as usual,
\begin{eqnarray}
 \langle O\rangle=
 \left\{{\mathcal Z}^{-1}\int_{\rm fields}O\ {\rm e}^{-S}\right\}_{
  \rho=\rho^{\prime}=0;\bar\rho=\bar\rho^{\prime}=0}.
\end{eqnarray}
$O$ denotes some gauge invariant functional
of the fields, possibly including the quark and antiquark
boundary fields $\zeta$ and $\bar\zeta$, which are obtained
by taking derivatives with respect to the quark boundary fields, viz.
\begin{eqnarray}
 \zeta(\mbox{\boldmath $\rm x$})\equiv P_{-}\zeta(\mbox{\boldmath $\rm x$})
 =\frac{\delta}{\delta\bar\rho(\mbox{\boldmath $\rm x$})}, \ \ \ \ \ 
 \bar\zeta(\mbox{\boldmath $\rm x$})\equiv \bar\zeta(\mbox{\boldmath
 $\rm x$})P_{-}
 =-\frac{\delta}{\delta\rho(\mbox{\boldmath $\rm x$})}.
\end{eqnarray}
The name ``Schr\"odinger functional'' derives from the
fact that such wave functionals arise naturally in
the Schr\"odinger representation of Quantum Field Theory~\cite{Symanzik:1981wd}, 
and the SF provides an example of a Quantum Field Theory 
defined on a manifold with a boundary.

Using correlation functions derived from the Schr\"odinger functional, 
it is possible to  define renormalised QCD parameters 
(the strong coupling and the quark masses),
as well as renormalised composite operators (e.g.~four-quark operators).
Such renormalization schemes based on the Schr\"odinger functional
(SF schemes) are attractive for the following reasons:
\begin{itemize}
\item The finite volume is part of the scheme definition, 
i.e.~all dimensionful quantities such as Euclidean time extent $T$, 
or boundary field parameters are scaled proportionally to $L$, 
the linear extent of the volume.
As a consequence $L$ remains the only scale in the system and can
be identified with the renormalization scale by setting $\mu=L^{-1}$.
Running parameters and operators then run with the 
size of the space-time volume, and one may apply recursive
finite size techniques to bridge large scale differences (cf.~subsect.~5.3.1)
\item
SF schemes are made quark mass independent
by imposing the renormalisation conditions in the chiral limit.
Fortunately, the SF boundary conditions introduce a gap in the spectrum
of the Dirac operator, which persists as the quark mass is 
taken to zero. This means that numerical simulations can
be performed in the chiral limit, and no chiral extrapolation is needed
to evaluate the renormalisation conditions.
\item
SF schemes are gauge invariant, no gauge fixing is needed.
\item
Perturbation theory up to two loops is still feasable, due to 
the existence of a unique absolute minimum of the action\cite{Luscher:1992an}.
This is to be contrasted with the situation on a hyper torus
where perturbation theory becomes very intricate already at the one-loop 
level.
\item A further technical advantage
consists in the possibility to use correlators involving
zero momentum  boundary quark and anti-quark fields.
This is convenient in perturbation theory, and it
leads to good numerical signals  and reduced cutoff effects as 
compared to gauge invariant correlators in a periodic setting.
\end{itemize}
All these nice properties come with a price: first of all,
the presence of the boundary means that even the pure gauge
theory suffers from O($a$) cutoff effects, caused by effective
local operators of dimension 4, such as
${\rm tr}\{F_{0k}F_{0k}\}$ and ${\rm tr}\{F_{kl}F_{kl}\}$,
integrated over the boundary. When the quarks are included, 
there is even a dimension 3 operator, 
which can be absorbed in a multiplicative rescaling of the
quark and antiquark boundary fields~\cite{Sint:1995rb}. 
At order $a$, one expects dimension 4 operators like 
$ \psibar\gamma_0 D_0\psi$  and $\psibar\gamma_k D_k\psi$
to contribute additional O($a$) effects~\cite{Luscher:1996sc}. It is important
to note that these cutoff effects are, unlike the O($a$) bulk
effects of Wilson quarks, not due to the breaking of a continuum
symmetry by the regularisation. Rather, such terms are to be
expected with any regularisation of the Schr\"odinger functional.
One may, however, write down a complete basis
of O($a$) counterterms which contribute to a given observable.
After reduction via the equations of motion, 
one typically ends up with 2-3 O($a$) boundary counterterms.
In practice it is then possible to monitor the size 
of the boundary O($a$) effects by varying the coefficients.
Perturbative results for these coefficients are 
often known to one-loop or even two-loop 
order~\cite{Bode:1999sm},
and a non-perturbative determination may be conceivable. 
In summary, with some extra work, the O($a$) boundary effects 
can be controlled and eventually eliminated. This is important,
as otherwise the SF renormalisation procedure risks to introduce O($a$) effects
even in  O($a$) improved regularisations such as tmQCD at maximal twist 
or lattice QCD with Ginsparg-Wilson quarks.

\subsection{Decoupling of heavy quarks in SF schemes}

Quark mass independent schemes are very convenient to 
study the scale evolution for a theory with fixed quark flavour content.
However, it also means that the decoupling of heavy quarks 
is not automatic, and one needs to match theories with different 
numbers of active flavours over quark thresholds.  
This is routinely done in perturbation theory, 
but it is not obvious that perturbation theory is adequate e.g.~for matching
the $\Nf=4$ and $\Nf=3$ effective theories over the charm quark threshold.
One possibility to study decoupling consists in 
introducing a quark mass dependent SF scheme which would allow
to study the  non-perturbative evolution over the quark threshold
until the heavy quark has decoupled. To define a mass dependent SF scheme 
it suffices to impose the renormalisation conditions at finite quark masses.
Unfortunately, it turns out that the decoupling of a 
heavy quark in such a scheme is only
{\em linear} in the inverse quark mass rather than quadratic.
If the quark decouples very slowly, this means that it has to
be kept longer in the evolution as an active degree of freedom,
which could mean that widely different scales have
to be accomodated on the same lattice.

\begin{figure}[ht]
\epsfxsize=10cm   
\epsfbox{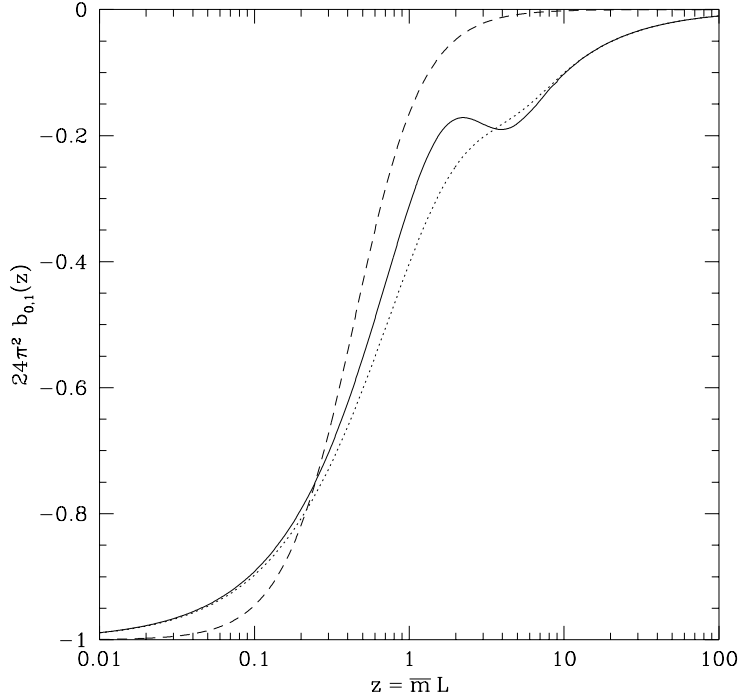}
\caption{Decoupling of a heavy quark in the one-loop $\beta$ function in two
SF schemes and in the MOM scheme. See text for further explanation \label{fig:decoupling}}
\end{figure}

An example from perturbation theory\cite{Sint:1995ch,Weisz:1995yz} is given in 
figure~\ref{fig:decoupling}.
It shows the one-loop $\beta$-function of the running coupling in 
the SF scheme as a function of $z=mL$, where $m$ is some renormalised 
quark mass (its precise definition is not required to one-loop order).
As $z=mL$ is varied from 0 to infinity, one expects to see a smoothed out
step function going from $-1$ to $0$ around the threshold $z=1$. 
The solid and dotted curves (from 2 different SF schemes) do 
indeed show this behaviour, but the
decoupling is rather slow compared to the MOM scheme\cite{Georgi:1976ve} (dashed line).

To understand this behaviour I propose a closer look at the Dirac operator
for free quarks and its spectrum in the continuum limit.

\subsubsection{Free quarks with SF boundary conditions}

Let us consider a free quark $\psi$ in the continuum with homogeneous SF
boundary conditions,
\begin{equation}
 P_{+}\psi(x)\mid_{x_0=0}=0, \qquad 
 P_{-}\psi(x)\mid_{x_0=T}=0. 
\end{equation}
Then $\gamma_5(\mbox{$\partial${\hspace{-2.1mm}/}}+m)$ is a hermitian
operator with smooth eigenfunctions and no zero modes~\cite{Sint:1993un}.
Evaluating the eigenvalue equation for any of its eigenfunctions $\varphi$ 
at the boundaries one finds,
\begin{eqnarray}
 P_{+}\gamma_5(\mbox{$\partial${\hspace{-2.1mm}/}}+m)\varphi\mid_{x_0=0}=0
 &\ \ \ \ \Rightarrow \ \ \ \ &
 (\partial_0-m)P_{-}\varphi\mid_{x_0=0}=0, \nonumber\\
 P_{-}\gamma_5(\mbox{$\partial${\hspace{-2.1mm}/}}+m)\varphi\mid_{x_0=T}=0
 &\Rightarrow&
 (\partial_0+m)P_{+}\varphi\mid_{x_0=T}=0.
\end{eqnarray}
The complementary components thus satisfy Neumann conditions modified by the
mass term $m$. The eigenvalues $\lambda$ are of the form 
$\lambda=\pm\sqrt{p_0^2+\mbox{\boldmath $p$}^2+m^2}$,
where $p_0$ is determined as non-vanishing solution of
$ \tan(p_0T)=-{p_0}/{m}$.
It is obvious from this equation that $p_0$ and thus $\lambda$
are not symmetric under $m\rightarrow -m$. This is generic
and can be understood as a consequence of chiral symmetry
breaking by the boundary conditions. 
As a result one expects, for any observable in the SF, the 
asymptotic small mass behaviour $\propto m$ (rather than $m^2$), 
and similarly for heavy quarks the corrections $\propto 1/m$
(instead of $1/m^2$), as illustrated in figure~\ref{fig:decoupling}.

At least for even numbers of flavours a possible 
way out consists in adding a twisted mass term and 
setting $m=0$. Then  
$\gamma_5\tau^1(\mbox{$\partial${\hspace{-2.1mm}/}}+i\muq\gamma_5\tau^3)$
is again hermitian. With this Dirac operator, the complementary field
components at the boundaries satisfy simple Neumann conditions,
and the spectrum is symmetric under a change of sign of 
the twisted mass term. A physically equivalent solution is 
obtained by staying with the
standard mass term and rotating the boundary projectors instead.
This will be discussed in more detail below. However, a caveat remains
as only the simultaneous decoupling of an even number of quarks
can be studied in this formalism. On the other hand, it may be sufficient 
to compare to perturbative decoupling in this slightly 
unphysical setting, in particular if a perturbative treatment 
turns out to be satisfactory.

\subsection{SF boundary conditions and chiral rotations}

Let us consider flavour doublets $\psi^{\prime}$ and $\psibar^{\prime}$
which satisfy homogeneous standard SF boundary conditions.
Performing a chiral rotation,
\begin{equation}
 \psi^{\prime}=\exp(i\alpha\gamma_5\tau^3/2)\psi, \qquad 
 \psibar^{\prime}=\psibar\exp(i\alpha\gamma_5\tau^3/2),
\end{equation}
one finds that the fields $\psi$ and $\psibar$ satisfy the chirally 
rotated boundary conditions,
\begin{eqnarray}
 P_{+}(\alpha)\psi(x)\mid_{x_0=0}=0, \ \ \ \ \ \ \ 
 P_{-}(\alpha)\psi(x)\mid_{x_0=T}=0, \nonumber\\
 \psibar(x)\gamma_0P_{-}(\alpha)\mid_{x_0=0}=0, \ \ \ \ \ \ \ 
 \psi(x)\gamma_0P_{+}(\alpha)\mid_{x_0=T}=0,
\end{eqnarray}
with the projectors,
\begin{eqnarray}
 P_{\pm}(\alpha)={\textstyle
  \frac{1}{2}}[1\pm\gamma_0\exp(i\alpha\gamma_5\tau^3)].
\end{eqnarray}
Special cases are $\alpha=0$ and $\alpha=\pi/2$ 
where one obtains,
\begin{equation}
 P_{\pm}(0)=P_{\pm}, \qquad  P_{\pm}(\pi/2)\equiv 
 Q_{\pm}={\textstyle\frac{1}{2}}(1\pm i\gamma_0\gamma_5\tau^3).
\end{equation} 
We perform again a change of variables in the functional integral.
Including mass terms as well, we label correlation functions
by a subscript $(m,\muq,P_+(\alpha))$, i.e.~we include the projector
defining the Dirichlet component of the quark field at
$x_0=0$. The generalisation of formula (\ref{eq:corr_identity})
then reads:
\begin{eqnarray}
 \langle O[\psi,\psibar]\rangle_{(m,\muq,P_{\pm})}=
 \langle O[R(\alpha)\psi,\psibar R(\alpha)]\rangle_{
        (\tilde{m},\tilde{\mu}_q,P_{\pm}(\alpha))},
\label{eq:equivSF}
\end{eqnarray}
with mass parameters $\tilde{m}$ and $\tilde{\mu_q}$ given by
\begin{equation}
 \tilde{m} =
 m\cos\alpha -\mu_q\sin\alpha, 
 \qquad
 \tilde{\mu_q} =
 m\sin\alpha +\mu_q\cos\alpha. 
\end{equation}
The boundary quark fields are included in this transformation
by replacing
\begin{eqnarray}
 \bar\zeta(\mbox{\boldmath $\rm x$}) \leftrightarrow
  \psibar(0,\mbox{\boldmath $\rm x$})P_{+}, \ \ \ \ \ 
 \zeta(\mbox{\boldmath $\rm x$}) \leftrightarrow
  P_{-}\psi(0,\mbox{\boldmath $\rm x$}).
\end{eqnarray}
This extends the equivalence between correlation functions
of tmQCD and standard QCD to correlation functions 
derived from the Schr\"odinger functional. 
Simple examples are provided by purely gluonic observables $O[U]$,
such as the SF coupling constant.
Eq.~(\ref{eq:equivSF}) then implies,
\begin{equation}
   \langle O[U]\rangle_{(0,\mu_{\rm R},P_{+})}=
   \langle O[U]\rangle_{(\mu_{\rm R},0,Q_{+})}.
\end{equation}
In other words, either the mass term is twisted 
and one stays with standard SF boundary conditions, or
the mass term is standard and the 
boundary conditions are fully twisted. In both cases one 
expects a quadratic dependence on the mass parameter and
hence a relatively fast decoupling of heavy quarks.

\subsection{SF schemes with Wilson quarks and $O(a)$ improvement}

\begin{figure}[ht]
\epsfxsize=9cm   
\epsfbox{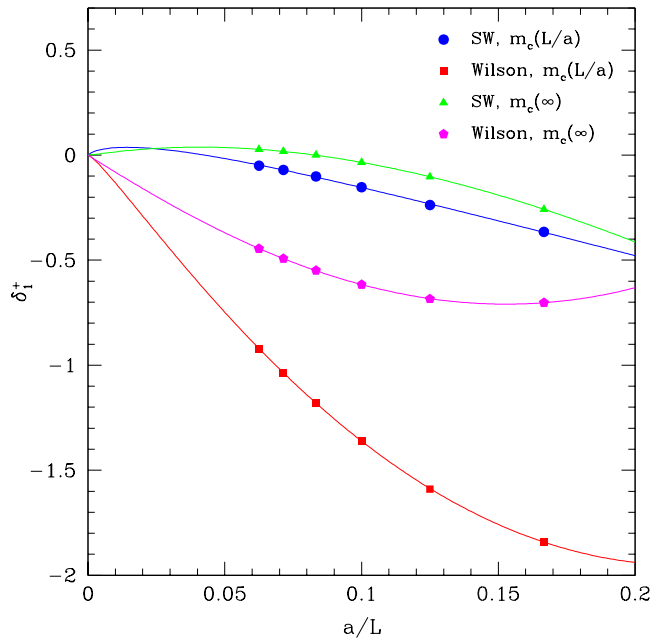}
\caption{Relative cutoff effects in the one-loop coefficient of 
the step-scaling function of the $B_K$ operator. Shown are
two different regularisations ($c_{\rm sw}=0,1$) with two definitions
of the zero mass point. \label{fig:cutoff}}
\end{figure}

From the discussion of O($a$) improvement in section~6
one may conclude that $\gamma_5\tau^1$-even observables
computed with Wilson quarks in a finite volume and with periodic
boundary conditions are automatically O($a$) improved at zero quark mass. 
As SF schemes are usually defined at zero quark mass, 
it seems natural to ask how the SF boundary conditions 
interfere with this property. It is useful to think 
of O($a$) effects to arise from different
sources. First there are the O($a$) boundary effects, which 
are cancelled by introducing the O($a$) boundary counterterms
to the action and the boundary quark and antiquark fields.
Second there are O($a$) effects from the bulk action which
may be cancelled by the Sheikholeslami-Wohlert term, and third there are the
O($a$) effects associated with  the composite operators in a given
correlation function. It is interesting to note that O($a$) cutoff effects
from the bulk action are often quite large in SF correlation functions. 
This is illustrated in figure \ref{fig:cutoff} which shows the
relative cutoff effects in the perturbative one-loop coefficient of 
the step-scaling function of the four-quark operator
needed for $B_K$~\cite{Palombi:2005zd}. The operator here is unimproved, and the 
boundary effects remain uncancelled in order to mimick
the non-perturbative procedure of~\cite{Guagnelli:2005zc}.
The most dramatic reduction of cutoff effects
occurs when the Sheikholeslami-Wohlert term is included. 
Moreover, this has the side effect to reduce the ambiguity in the zero mass point, 
so that with the standard SF it makes sense to 
implement O($a$) improvement even if it is not complete.

\subsection{The Schr\"{o}dinger functional and $O(a)$ improvement}

The reason why automatic $O(a)$ improvement fails is that 
the $\gamma_5\tau^1$-transformation (\ref{eq:gam5tau1}) 
changes the projectors of the quark boundary conditions,
\begin{eqnarray}
 P_{\pm}\gamma_5\tau^1=\gamma_5\tau^1 P_{\mp}.
\end{eqnarray}
The boundary conditions, just like mass terms, 
define a direction in chiral flavour space. This means that the
$\gamma_5\tau^1$-transformation yields inequivalent correlation
functions even in the chiral limit. For a $\gamma_5\tau^1$-even
operator $O$ one finds
\begin{eqnarray}
 \langle O\rangle_{(m,\muq,P_{+})}\ \rightarrow \ 
  \langle O\rangle_{(-m,\muq,P_{-})}.
\end{eqnarray}
It thus appears that the standard SF does not allow for
the definition of $\gamma_5\tau^1$-even correlation functions,
and bulk O($a$) improvement is not automatic.

A possible solution is obtained by changing the projectors 
used to specify the Dirichlet components such
that they commute with $\gamma_5\tau^1$. Allowing for an additional 
flavour structure one may think 
of ${\textstyle\frac{1}{2}}(1\pm\gamma_0\tau^3)$ or
\begin{equation}
 Q_{\pm}={\textstyle\frac{1}{2}}(1\pm i\gamma_0\gamma_5\tau^3).
\end{equation}
Interestingly, the projectors $Q_\pm$ also appear in the
chiral rotation of the SF by $\alpha=\pi/2$. Besides automatic
O($a$) improvement, the implementation of these boundary conditions
may lead to some interesting checks of universality by comparing
SF correlation functions in the standard framework and
at maximal twist. Note that this direct comparison was not
possible in \cite{Frezzotti:2001ea,DellaMorte:2001ys}, where a twisted mass term
was introduced whilst keeping the standard SF boundary conditions.

\subsection{The SF with chirally rotated boundary conditions}

The implementation of some given boundary
conditions is not straightforward on the lattice,
and some care has to be taken to ensure that one really
ends up with the desired continuum theory.
A successful implementation of the maximally twisted boundary conditions
involving the projectors $Q_\pm$ has been described in \cite{Sint:2005qz},
and relies on an orbifold construction to ensure the correct continuum 
limit.

\subsubsection{Symmetries and Counterterms}

Apart from the absence of a dimensionful parameter, the symmetries of 
the SF with maximally twisted boundary conditions 
are identical to those of tmQCD. One may then list the 
possible boundary counterterms of dimension 3 allowed by the 
symmetries:
\begin{equation}
 K_1=\psibar i\gamma_5\tau^3\psi, \qquad
 K_\pm =\psibar Q_\pm \psi.  
\end{equation}
As time reflection combined with a flavour permutation is a symmetry
of the SF, it is enough to discuss the counterterms at $x_0=0$.
$K_1$ corresponds to the logarithmically divergent boundary counterterm
in the standard SF, which leads to a multiplicative renormalization
of the quark boundary fields.  
The operator $K_{+}$ only involves Dirichlet
components at $x_0=0$ and is therefore irrelevant for most
correlation functions used in practice. The remaining operator
$K_-$  only contains non-Dirichlet boundary components. 
If rotated back to the primed basis it 
becomes proportional to  $\psibar^{\prime}i\gamma_5\tau^3P_{-}\psi^{\prime}$,
which violates flavour symmetry and parity just like a twisted mass term.
As these are symmetries which are restored in the continuum limit 
one concludes that this counterterm must be scale-independent. Its coefficient
can be fixed by requiring that a parity violating SF correlation function
vanishes at finite $a$.

This analysis can be extended to dimension 4 operators
which appear as  O($a$) boundary counterterms~\cite{ssint}. 
It turns out that the situation is comparable to the standard SF, 
i.e.~there are a couple of counterterms
which one needs to tune in order to eliminate the O($a$) boundary artefacts.

\subsection{An example from perturbation theory}

In perturbation theory, the values of all boundary counterterms
are known, so that one may study both the equality of 
properly matched standard and twisted SF correlation functions,
and confirm automatic bulk O($a$) improvement.
A first example is given by the SF coupling, which 
can be related perturbatively to the $\overline{\rm MS}$-coupling,
\begin{eqnarray}
 \bar{g}^2(L)=g_{\overline{\rm MS}}^2(\mu)+
              k_1(\mu L)g_{\overline{\rm MS}}^4(\mu)+O(g^6).
\end{eqnarray}
The fermionic contribution to the one-loop coefficient,
$k_1 =k_{1,0}+N_{\rm f}k_{1,1}$, has been computed in~\cite{Sint:1995ch}, yielding
$ k_{1,1}=-0.039863(2)/(4\pi)$.
In practice, the perturbative data is obtained for a sequence of lattices,
and one then expects the asymptotic large $L/a$ behaviour:
\begin{eqnarray}
 f(L/a) \sim r_0+(a/L)[r_1+s_1 \ln(a/L)]+O(a^2).
\end{eqnarray}
Here $r_0=k_{1,1}$ is the continuum limit value, and the O($a$) effects
lead to non-vanishing values of $r_1$ and $s_1$. In the standard Schr\"odinger
functional set-up one expects that $r_1$ is eliminated by the
boundary counterterm proportional to ${\rm tr}(F_{0k}F_{0k})$, whereas
$s_1$ is due to bulk O($a$) effects from the action, and thus
proportional to $c_{\rm sw}^{(0)}-1$. 
On the other hand, with twisted SF boundary conditions one
expects that $r_0$ remains the same, due to universality, $r_1$
is cancelled again by a boundary counterterm, and $s_1$ should vanish
independently of the value of $c_{\rm sw}^{(0)}$.This expectation is indeed 
confirmed numerically.
A similar test can be performed with the tree level quark propagator
in a non-vanishing gauge background field, induced by
choosing non-vanishing gauge field boundary values $C_k^{}$ and $C_k'$.
One then expects that, with the correct tree-level boundary 
counterterms, the bulk O($a$) lattice artefacts will again be either proportional
to $c_{\rm sw}^{(0)}-1$ (standard SF) or absent (twisted SF).
Again this expectation is confirmed. 
However, in contrast to the SF coupling this test can not be 
extended beyond the tree level, unless one fixes the gauge.

\section{Conclusions}

Lattice QCD with Wilson type quarks remains an attractive 
regularisation of lattice QCD. Some of its problems can
be alleviated by introducing a chirally twisted quark mass term.
While the theories remain equivalent in the continuum limit,
the twisted mass term supplies an infrared 
bound on the spectrum of the Wilson-Dirac operator  
which renders the quenched and partially quenched approximations 
well-defined. Some of the  notorious lattice renormalisation 
problems of standard Wilson quarks can be by-passed, and
tmQCD at maximal twist is automatically O($a$) improved.
These advantages are balanced by parity and flavour breaking
and the fact that tmQCD comes naturally with an even number of quarks.

The Schr\"odinger functional has become an indispensable tool
to tackle non-perturbative renormalisation problems in lattice QCD 
However, the standard set-up leads to a slow decoupling of
heavy quarks, and is in conflict with automatic 
O($a$) improvement of massless Wilson quarks. 
This motivates the application of a chiral twist 
to the SF boundary conditions. It is thus possible
to extend equivalence between tmQCD and standard QCD to
correlation functions derived from the Schr\"odinger functional.
This allows for interesting tests of universality and
the maximally twisted SF is compatible with automatic
O($a$) improvement, as I have illustrated with simple perturbative examples.

\section*{Acknowledgments}
I would like to thank the organisers of the workshop, in particular
Yoshinobu Kuramashi for the invitation to this very pleasant meeting
in the beautiful city of Nara. Support by the Japanese government
and by the Spanish government through a ``Ram{\'o}n y Cajal'' fellowship 
is gratefully acknowledged.

\end{document}